\documentclass[showpacs,aps,prc,longbibliography,nofootinbib,showkeys,superscriptaddress,twocolumn]{revtex4-1}

\usepackage[colorlinks, urlcolor=blue,linkcolor=blue, anchorcolor=blue, citecolor=blue]{hyperref}
\usepackage[utf8]{inputenc}
\usepackage{natbib}
\usepackage{graphicx}
\usepackage{xcolor}
\usepackage{amsmath}
\usepackage{bm}
\usepackage{multirow}
\usepackage{array}
\usepackage{amsthm}
\usepackage{mathtools}
\usepackage{physics}
\usepackage{xcolor}
\usepackage{graphicx}
\usepackage{adjustbox}
\usepackage{placeins}
\usepackage[T1]{fontenc}
\usepackage{lipsum}
\usepackage{csquotes}

\newcommand{\music}{{\sc MUSIC}}
\newcommand{\isd}{{\sc iS3D}}
\newcommand{\iss}{{\sc iSS}}
\newcommand{\urqmd}{{\sc UrQMD}}
\newcommand{\snn}{\sqrt{s_\mathrm{NN}}}

\newcommand{\yb}{y_\mathrm{b}}
\newcommand{\Npart}{N_\mathrm{part}}

\graphicspath{{figs}}

\begin{document}

\title{Bulk medium properties of heavy-ion collisions from the beam energy scan\\with a multistage hydrodynamic model}

\author{Lipei Du}
\affiliation{Department of Physics, McGill University, Montreal, Quebec H3A 2T8, Canada}

\date{\today}

\begin{abstract}
    I introduce a method to reconstruct full rapidity distributions of charged particle multiplicity and net proton yields, crucial for constraining the longitudinal dynamics of nuclear matter created in the beam energy scan program. Employing rapidity distributions within a multistage hydrodynamic model calibrated for Au+Au collisions at $\snn=7.7-200\,$GeV, I estimate the total energy and baryon number deposited into the collision fireball, offering insights into initial dynamics and the identification of nuclear remnants. I explore the potential of rapidity-dependent measurements in probing equations of state at finite chemical potentials. Furthermore, I compare the freeze-out parameters derived from both hydrodynamics and thermal models, highlighting that the parameters extracted via thermal models represent averaged properties across rapidities.
\end{abstract}

\maketitle

\section{Introduction}\label{sec:intro}

Investigations of relativistic heavy-ion collision across various beam energies, such as those in the beam energy scan (BES) program, stand as primary methods for mapping the QCD phase diagram \cite{Bzdak:2019pkr,Sorensen:2023zkk,Du:2024wjm}. Given the highly dynamic nature of the collision fireball, it is crucial to perform model-to-data comparisons to derive medium properties from experimental measurements. Employing multistage hydrodynamic models that consist of diverse physics has become the standard approach for modeling these collisions which go through various phases \cite{Heinz:2013th,Shen:2014vra}. These models have achieved a great success in describing nuclear collisions at the top RHIC and LHC energies, assuming a charge-neutral quark-gluon plasma (QGP) \cite{Bernhard:2019bmu,JETSCAPE:2020mzn,JETSCAPE:2020shq,Nijs:2020ors}. In these scenarios, the longitudinal expansion is often approximated as boost-invariant, with calculations conducted in a (2+1)-dimensional setup, omitting consideration for conserved charges.

At lower beam energies, particularly at the tens of GeV center-of-mass energies and below, the nuclear matter can attain high charge densities and exhibit significant departure from boost-invariance. The substantial thermodynamic variations and absence of boost-invariance in the beam direction underscore the importance of comprehending longitudinal evolution, involving nontrivial longitudinal flow and charge transport \cite{Denicol:2018wdp,Du:2021zqz,Du:2021uxo,Du:2023gnv}. This necessitates a comprehensive (3+1)-dimensional dynamic modeling, including the evolution of conserved charges such as baryon number, electric charge, and strangeness \cite{Fotakis:2022usk}. Calibration of (3+1)-dimensional dynamics crucially relies on rapidity-dependent experimental measurements. For instance, the charge particle multiplicity offers insights into constraining entropy and energy densities, while net proton yields aid in constraining net baryon density along the beam direction \cite{Denicol:2018wdp,Du:2022yok}.

Recent developments in the (3+1)-dimensional multistage framework have attained substantial progress. These advancements encompass rapidity-dependent initial conditions \cite{Bozek:2010bi,Bozek:2011ua,Shen:2017bsr,Denicol:2018wdp,Shen:2022oyg,Du:2022yok,De:2022yxq}, incorporation of charge evolution in hydrodynamic simulations \cite{Denicol:2018wdp,Du:2019obx}, computation of equations of state (EoS) \cite{Monnai:2019hkn,Noronha-Hostler:2019ayj,Monnai:2021kgu,Kapusta:2021oco} and transport coefficients \cite{Greif:2017byw} at finite chemical potentials, and enhancements in particlization samplers to accommodate multiple charges \cite{McNelis:2019auj,Oliinychenko:2019zfk}. These advancements have yielded invaluable insights into nuclear matter properties at finite chemical potentials on various aspects, including initial baryon and energy deposition \cite{Shen:2017bsr,Du:2018mpf,Shen:2022oyg,Du:2022yok}, longitudinal charge transport phenomena \cite{Denicol:2018wdp,Du:2021zqz,Du:2021uxo}, and the EoS at finite chemical potentials \cite{Monnai:2019hkn,Shen:2022oyg}. Furthermore, the inclusion of rapidity-dependence has motivated rapidity scan approaches to explore the QCD phase diagram \cite{Karpenko:2018xam,Begun:2018efg,Du:2023gnv} and potentially aid in searching for the QCD critical point \cite{Brewer:2018abr,Du:2021zqz,Li:2023kja} through rapidity measurements.

Despite the theoretical advancements, systematically calibrating a comprehensive (3+1)-dimensional framework to understand aspects like initial baryon stopping \cite{Shen:2017bsr,Shen:2022oyg,Shen:2023awv} and charge transport properties \cite{Denicol:2018wdp,Du:2021zqz} encounters challenges due to limited rapidity measurements across varying beam energies. Presently, experimental measurements predominantly focus on the midrapidity region within a finite rapidity window, constrained by detector design and statistical considerations \cite{STAR:2008med,STAR:2017sal}. Moreover, observables are measured around midrapidity within fixed rapidity windows across beam energies, and non-monotonic behaviors in their energy dependence often hint at intriguing phenomena. However, at lower beam energies with smaller beam rapidities, the system's thermodynamic properties exhibit more pronounced variations along rapidity \cite{Du:2023gnv}. This suggests that observables within fixed rapidity windows encompass averaging over thermodynamic properties that undergo more dramatic changes at lower beam energies. Consequently, interpreting results across beam energies becomes more complex, particularly when searching for the critical point \cite{Vovchenko:2021kxx,Li:2023kja}.

%
\begin{table*}[!bpht]
\centering
\caption{Measurements of identified particle yields and charged particle multiplicities observed in Au+Au and Pb+Pb collisions across different beam energies. Adapted from Ref.~\cite{Du:2023gnv}, with additional data included.}
\vspace{1mm}
\begin{tabular}{cccc}
\hline
\hline
$\snn$ & Collision system & Observable & Collaboration  \\ 
    \hline
    7.7 GeV & Au+Au & $dN/dy|_{|y|<0.1}$ of $\pi^+$, $K^+$, $p$ and $\bar{p}$ \cite{STAR:2017sal} & STAR\\
    \hline
    \multirow{2}{*}{8.8 GeV}  & \multirow{2}{*}{Pb+Pb} & $dN/dy$ of $p$ and $\bar{p}$ \cite{NA49:2010lhg} and of $\pi^+$ and $K^+$ \cite{NA49:2012rsi}  & NA49 \\
    & & $dN/d\eta$ of charged particles \cite{NA50:2002edr} & NA50\\
    \hline
    17.3 GeV & Pb+Pb & $dN/dy$ of $p-\bar{p}$ \cite{NA49:1998gaz} and of $\pi^+$ and $K^+$  \cite{NA49:2012rsi} & NA49\\
    \hline
    \multirow{2}{*}{19.6 GeV} & \multirow{2}{*}{Au+Au} & $dN/d\eta$ of charged particles \cite{Back:2002wb} &  PHOBOS\\ 
    & & $dN/dy|_{|y|<0.1}$ and $p_T$ spectra of $\pi^+$, $K^+$, $p$ and $\bar{p}$ \cite{STAR:2017sal} & STAR\\
    \hline
    \multirow{3}{*}{62.4 GeV} & \multirow{3}{*}{Au+Au} & $dN/d\eta$ of charged particles \cite{PHOBOS:2005zhy} &  PHOBOS\\ 
    & & $dN/dy$ of $\pi^+$ and $K^+$ \cite{BRAHMS:2009acd}, and of $p$ and $\bar{p}$ \cite{BRAHMS:2009wlg}& BRAHMS\\
    & & $dN/dy|_{|y|<0.1}$ of $\pi^+$, $K^+$, $p$ and $\bar{p}$ \cite{STAR:2008med} & STAR\\
    \hline
    \multirow{3}{*}{130 GeV} & \multirow{3}{*}{Au+Au} &  $N^{\bar p}/N^{p}$ around midrapidity  \cite{PHOBOS:2001rob, STAR:2001rbj} & PHOBOS, STAR\\
    & & $(N^{\bar p}/N^{p})(y)$ \cite{BRAHMS:2001kqb}, $dN^{\bar p}/dy$ and $dN^{p}/dy$ \cite{STAR:2003ryp} & BRAHMS, STAR\\
    & & $dN^{\bar p}/dy$ and $dN^{p}/dy$ around midrapidity \cite{PHENIX:2001vgc} & PHENIX\\
    \hline
    \multirow{4}{*}{200 GeV} & \multirow{4}{*}{Au+Au} & $dN/d\eta$ of charged particles \cite{Back:2002wb} &  PHOBOS\\ 
    & & $dN/dy$ of $\pi^+$ and $K^+$ \cite{BRAHMS:2004dwr}, and of $p$ and $\bar{p}$ \cite{BRAHMS:2003wwg} & \multirow{2}{*}{BRAHMS}\\
    & & $dN/d\eta$ of charged particles \cite{BRAHMS:2001llo}& \\
    & & $dN/dy|_{|y|<0.1}$ of $\pi^+$, $K^+$, $p$ and $\bar{p}$ \cite{STAR:2008med} & STAR\\
\hline
\hline
\end{tabular}

\label{tab:data}
\end{table*}

In this study, I utilize available rapidity-dependent measurements, specifically focusing on the charged particle multiplicity and net proton yields, to investigate their universal scaling properties across various beam energies. With these identified universal distributions, I aim to construct complete rapidity distributions for beam energies lacking direct measurements (Sec.~\ref{sec:models}). These reconstructed distributions offer a means to calibrate the bulk dynamics, especially in the longitudinal direction, for collisions at BES (Sec.~\ref{sec:reconst_27}). Moreover, utilizing this calibrated framework allows for the calculation of the total net baryon number and total energy deposited within the collision fireball. This analysis aids in probing the mechanisms behind initial baryon and energy deposition and in identifying the nuclear remnants in the fragmentation region and potentially string junction (Sec.~\ref{sec:fragment}). I investigate the constraints imposed by rapidity-dependent yields on the EoS at finite chemical potentials (Sec.~\ref{sec:eos}). I study the thermodynamic properties on the hydrodynamic freeze-out hypersurface, aiming to facilitate interpreting freeze-out parameters derived from thermal models (Sec.~\ref{sec:freezeout}). Lastly, I summarize the key insights obtained from this study (Sec.~\ref{sec:conclusions}).

\section{Model and setup}\label{sec:models}

\subsection{Multistage hydrodynamic model}\label{sec:hybrid_setup}

I utilize a (3+1)-dimensional multistage hydrodynamic framework with parametric initial conditions \cite{Denicol:2018wdp,Du:2022yok} to simulate Au+Au collisions at $\snn=7.7,\,\,$$19.6,\,\,$$27,\,\,$$39,\,\,$$54.4,\,\,$$62.4,\,\,$$130,\,\,$$200\,$ GeV. The approach involves constructing initial entropy and baryon densities by extending the nucleus thickness function with parametrized longitudinal profiles, following the method described in Ref.~\cite{Denicol:2018wdp}. The hydrodynamic stage initiates at a constant proper time $\tau_0$ with Bjorken flow \cite{Bjorken:1982qr}. The evolution of the energy-momentum tensor and net baryon current, simulated using \music{} \cite{Schenke:2010nt,Schenke:2011bn,Paquet:2015lta},  considers dissipative effects from the shear stress tensor and net baryon diffusion current, with the bulk viscous pressure being excluded in this work. For the simulations, I use a specific shear viscosity $\eta/s$ that exhibits dependencies on both temperature $T$ and baryon chemical potential $\mu_B$ \cite{Shen:2020jwv}, alongside a baryon diffusion coefficient $\kappa$ derived from the Boltzmann equation in the relaxation time approximation at the massless limit \cite{Denicol:2018wdp}
\begin{equation}
    \kappa = \frac{C_B}{T} n \left(\frac{1}{3} \coth\left(\frac{\mu_B}{T}\right) - \frac{n T}{e + p}\right)\,,
\end{equation}
with free parameter $C_B=0.3$. Here $e, n, p$ are energy density, baryon density, and pressure, respectively.

I adopt an equation of state, referred to as ``NEOS'' from Ref.~\cite{Monnai:2019hkn}, which is constructed by a smooth interpolation between high-temperature results derived from lattice QCD EoS and low-temperature EoS of a hadron resonance gas. The lattice QCD EoS is expanded to finite baryon chemical potential through the Taylor expansion method \cite{HotQCD:2014kol,HotQCD:2012fhj,Ding:2015fca}. When the system expands and cools, I implement the particlization process on a freeze-out hypersurface defined by a constant freeze-out energy density at $e_\mathrm{fo}{\,=\,}0.35$ GeV/fm$^3$, which aligns with the chemical freeze-out line extracted by the STAR Collaboration \cite{STAR:2017sal}. To sample hadrons on this freeze-out surface, I utilize the \isd{} \cite{McNelis:2019auj} and \iss{} \cite{Shen:2014vra} particle samplers,\footnote{%
Both particle samplers are utilized in this study, as each has its distinct advantages. The excellent agreement between these two samplers has been demonstrated in Ref.~\cite{Du:2023gnv}.
}
applying the Cooper-Frye prescription \cite{Cooper:1974mv} and considering off-equilibrium effects from shear stress and baryon diffusion. The validation of the \isd{} sampler includes comparison with smooth distributions from the Cooper-Frye prescription (detailed in Appendix~\ref{app:is3d}) and results obtained using \iss{}.
For the hadronic afterburner, I employ \urqmd{} \cite{Bass:1998ca,Bleicher:1999xi}. I account for weak decay feed-down contributions in the net proton yields when comparing them to experimental data.

\subsection{Reconstructing rapidity distributions}\label{sec:rap_reconst}

\begin{figure}[!bt]
\begin{center}
\includegraphics[width= 0.96\linewidth]{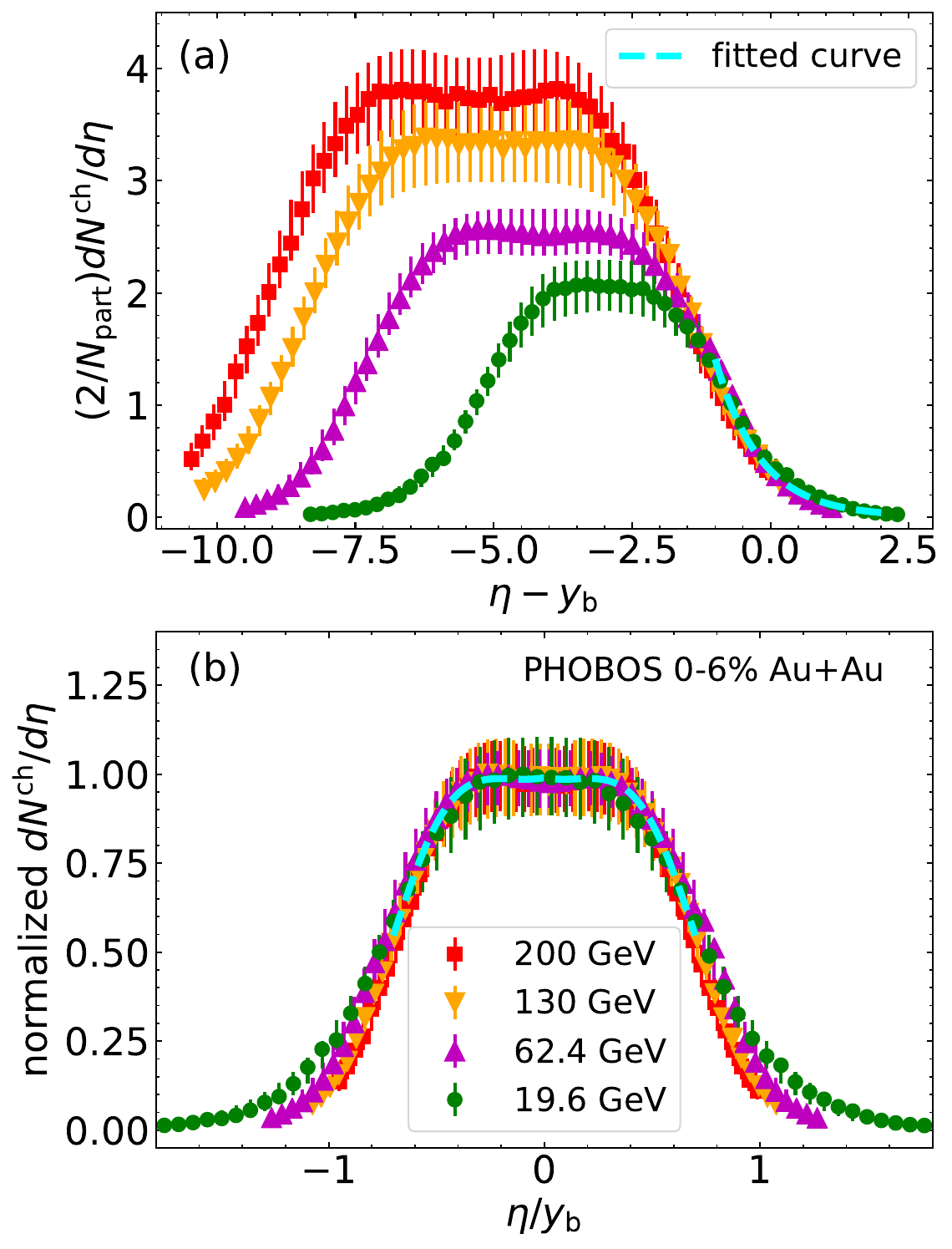}
    \caption{%
        (a) The distribution of charge particle multiplicity per participant pair in the shifted pseudo-rapidity $\eta-y_b$, and (b) the distribution of normalized charge particle multiplicity in $\eta/y_b$. Markers represent experimental measurements, while dashed curves illustrate fitted curves corresponding to (a) the fragmentation region and (b) the central plateau region.
        }
    \label{fig:charged_scaling}
\end{center}
\end{figure}

Rapidity-dependent measurements play a crucial role in constraining theoretical models for collisions at the beam energy scan. Notably, the charged particle multiplicity in pseudo-rapidity, $dN^\mathrm{ch}/d\eta$, holds significance in probing longitudinal distributions of entropy and energy densities, while the rapidity-dependent net proton yields, $dN^{p-\bar p}/dy$, aid in constraining the distribution of net baryon density. Regrettably, contemporary measurements have predominantly focused on the midrapidity region, resulting in limited rapidity-dependent data, largely obtained from early experiments. In this section, leveraging available rapidity data across diverse beam energies,\footnote{%
Some measurements at various beam energies are presented in Table~\ref{tab:data} for the convenience of readers.
}
I summarize existing methods and propose additional ones to discern universal scaling properties among these data points. By fitting these points with universal curves, I establish an approach to reconstruct rapidity-dependent distributions that are not accessible experimentally. These reconstructed distributions offer useful tools for calibrating theoretical models.\footnote{%
In this work, I focus on symmetric nucleus-nucleus collisions. Although similar investigations could be conducted for asymmetric small collision systems, they are beyond the scope of the current study.
}

\begin{figure}[!tb]
\begin{center}
\includegraphics[width= 0.9\linewidth]{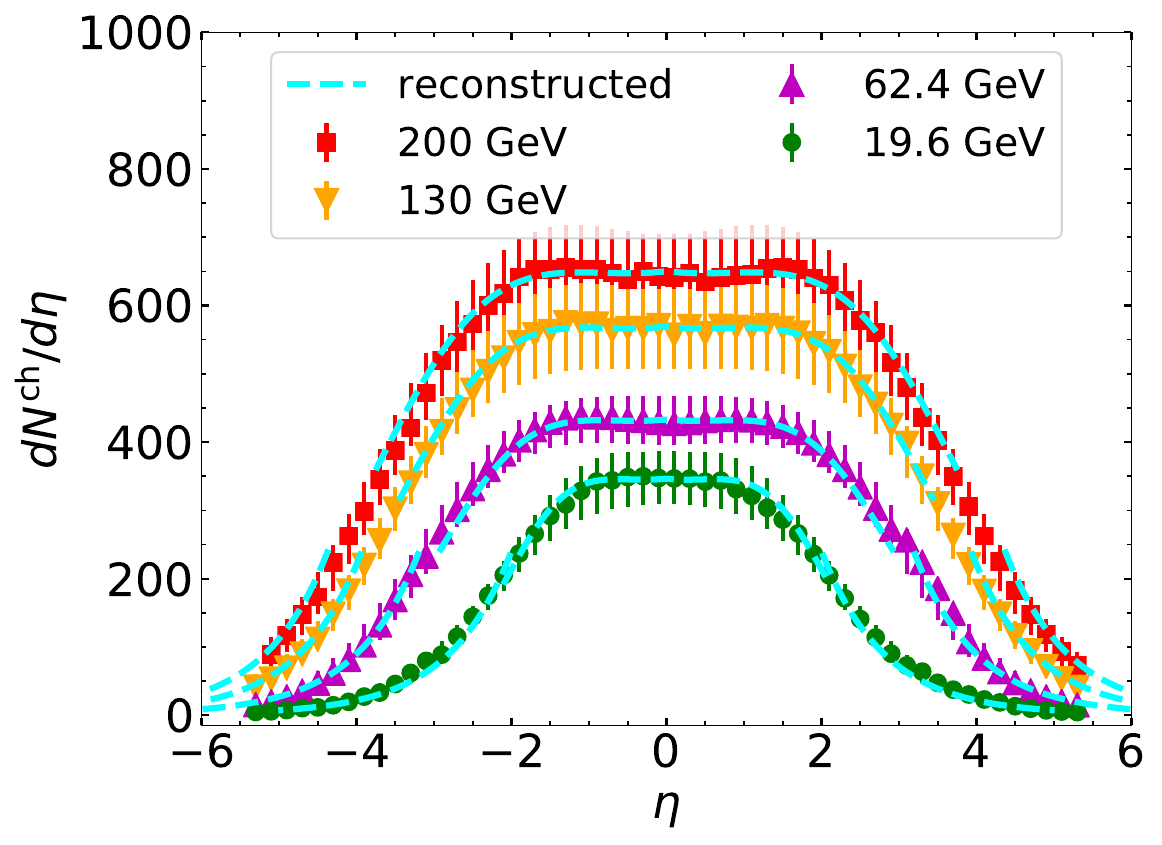}
    \caption{%
        The distribution of charge particle multiplicity in pseudo-rapidity $\eta$. Experimental measurements for 0-6\% Au+Au collisions are represented by markers, while the dashed curves depict the reconstructed rapidity distribution derived from the fitted curves obtained in Fig.~\ref{fig:charged_scaling}.
        }
    \label{fig:charged_reconst}
\end{center}
\end{figure}

I take advantage of the PHOBOS measurements for $dN^\mathrm{ch}/d\eta$ with extensive pseudo-rapidity coverage \cite{Back:2002wb}, and identify universal properties within distinct regions: the central plateau and the tail in fragmentation region. The well-established phenomenon of limiting fragmentation \cite{Benecke:1969sh} for the forward tail becomes evident across various energies in the rest frame of one of the colliding nuclei. Illustrated in Fig.~\ref{fig:charged_scaling}(a), the distribution of charged particle multiplicity per participant pair against the shifted pseudo-rapidity $\eta-\yb$ shows energy-independent distributions in the fragmentation region. For the range where $\eta-\yb>-1$, an exponential function with an offset, $a \exp(-b x) + c$, with the optimal fit parameters $a=0.41$, $b=1.23$, and $c=0.007$, describe the data well. 
To analyze the central plateau region, I normalize $dN^\mathrm{ch}/d\eta$ relative to its value at midrapidity and then plot it against the rescaled $\eta/\yb$ in Fig.~\ref{fig:charged_scaling}(b). Remarkably, the resulting distributions from central Au+Au collisions, spanning a tenfold range in collision energy, converge onto a universal curve around the plateau region. Within the region where $|\eta/\yb|<0.7$, I employ a fitting function \cite{PHOBOS:2010eyu}:
\begin{equation}
    F(x)=\frac{c \sqrt{1 - 1 /(a \cosh x)^2}}{1 + \mathrm{e}^{(|x| - b) / \delta}}\,,
\end{equation}
where the best-fit parameters are determined to be $c = 1.79$, $a = 1.21$, $b = 0.64$, and $\delta = 0.16$. The universal fitting is utilized for this region where the data from various beam energies collapse, as shown in Fig.~\ref{fig:charged_scaling}(b).

Utilizing the fitting curves obtained for both the fragmentation region and the central plateau region, reconstructing $dN^\mathrm{ch}/d\eta$ becomes straightforward given the participant number $N_\mathrm{part}$ and the charged particle multiplicity around midrapidity. Estimating the former involves employing the Glauber model \cite{Miller:2007ri}, while the latter can be derived using the following relationship \cite{PHOBOS:2010eyu}:
\begin{equation}\label{eq:mid_charge}
\frac{2}{N_\mathrm{part}}\left.\frac{dN^\mathrm{ch}}{d\eta}\right|_{\eta=0}=0.77\, s_{_\mathrm{NN}}^{0.15}\,,
\end{equation}
where $(dN^\mathrm{ch}/d\eta)|_{\eta=0}$ represents the midrapidity multiplicity, and $s_{_\mathrm{NN}}$ denotes the beam energy. Leveraging these quantities, I reconstruct the full distributions and validate them against experimental measurements. As depicted in Fig.~\ref{fig:charged_reconst}, the reconstructed distributions exhibit remarkable agreement with the experimental data. To reconstruct the charged particle multiplicity, I utilized $\Npart$ values of 337, 340, 340, and 344 for central Au+Au collisions at 19.6, 62.4, 130, and 200 GeV, respectively, as reported in Refs.~\cite{Back:2002wb,PHOBOS:2005zhy}.

\begin{figure}[!tb]
\begin{center}
\includegraphics[width= 0.93\linewidth]{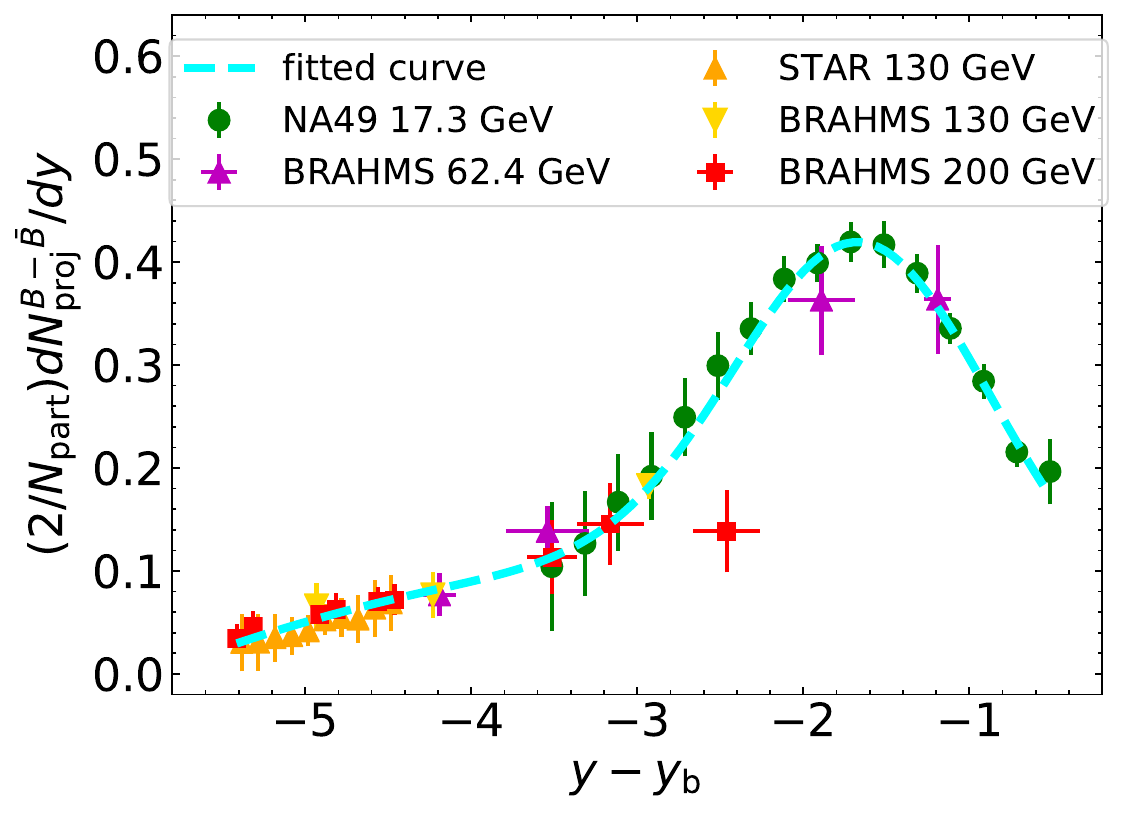}
    \caption{%
        The net baryon rapidity density attributed to the projectile after subtraction of the target contribution. Experimental estimates are denoted by markers, while the dashed curve illustrates a fitted curve comprising a Gaussian distribution and a polynomial distribution sum.
        }
    \label{fig:baryon_scaling}
\end{center}
\end{figure}

The observed double-humped structure in net proton distributions is commonly associated with the incoming nucleons from the projectile or target following rapidity loss. Deriving net baryon distributions from net proton distributions and subsequently subtracting the target contribution, universal scaling properties have been revealed \cite{BRAHMS:2009wlg,Braun-Munzinger:2020jbk,Hoelck:2023jme} in the resulting distributions when plotted against shifted rapidity ($y-\yb$). In this approach, the target contribution is parametrized as the average of two exponential functions,\footnote{%
Regarding the potential relationship of this parametrization to the gluon junctions, see Sec.~\ref{sec:fragment} for a discussion.
}
\begin{equation}\label{eq:targ}
    f_\mathrm{targ}(x)=C[\exp(-x)+\exp(-x/2)]/2\,,
\end{equation}
where $x$ represents shifted rapidity ($y-\yb$), and $C$ serves as a constant prefactor \cite{BRAHMS:2009wlg,Braun-Munzinger:2020jbk}. The determination of the prefactor $C$ often involves symmetry considerations, ensuring equal contributions from the projectile and target at mid-rapidity. Additionally, estimating the net baryon distribution from the net proton one usually assums that the net baryon yield in the entire phase space, i.e., the rapidity-integrated net baryon number, is equal to the number of participants (wounded nucleons) \cite{Braun-Munzinger:2020jbk}.

Here, I aim to reproduce the observed universal scaling reported in Refs.~\cite{BRAHMS:2009wlg,Braun-Munzinger:2020jbk} using data from Pb+Pb collisions at 17.3 GeV and Au+Au collisions at 62.4 and 200 GeV. Subsequently, I extend the analysis by including data from Pb+Pb collisions at 8.8 GeV and Au+Au collisions at 130 GeV, examining whether these additional datasets exhibit similar universal behaviors. The rapidity density of protons at 8.8 GeV is taken from Ref.~\cite{NA49:2010lhg}, reporting participant numbers as $\Npart=356\pm1$ for 0-5\% and $\Npart=292\pm2$ for 5-12.5\% collisions. For the 0-5\% collisions, I adopt the $\Npart=352$ estimated in Ref.~\cite{Braun-Munzinger:2020jbk}. The net baryon distribution for Pb+Pb collisions at 17.3 GeV is estimated from net proton measurements by the NA49 Collaboration \cite{NA49:1998gaz}, associating it with $\Npart=352\pm12$. Additionally, Ref.~\cite{Braun-Munzinger:2020jbk} presents an estimation of the net baryon distribution derived from the net proton distribution, utilizing $\Npart=352$.

\begin{figure}[!tb]
\begin{center}
\includegraphics[width= 0.9\linewidth]{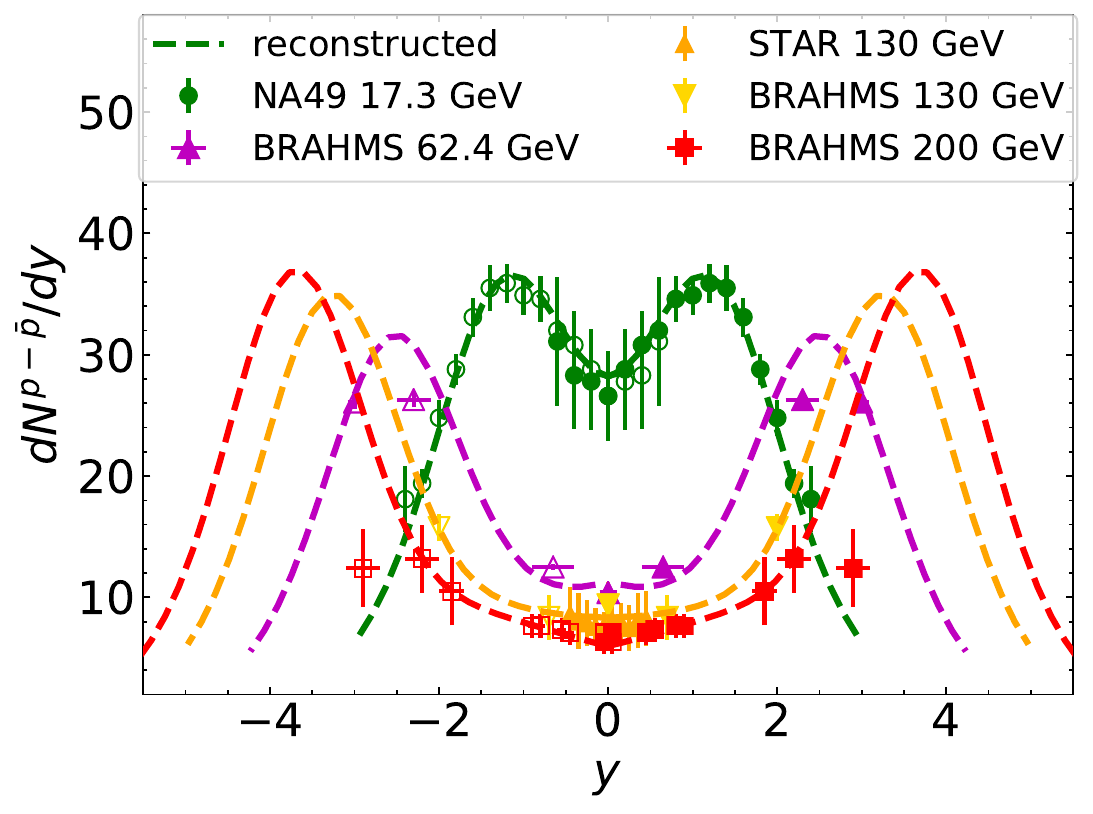}
    \caption{%
        The net proton rapidity density across various beam energies. Experimental measurements are denoted by markers, while the dashed curves illustrate the reconstructed rapidity distribution derived from the fitted curve obtained in Fig~\ref{fig:baryon_scaling}. 
        }
    \label{fig:net_proton_reconst}
\end{center}
\end{figure}

The net baryon distributions per participant pair at 62.4 and 200 GeV are reported by the BRAHMS Collaboration in Ref.~\cite{BRAHMS:2009wlg}. For 0-10\% collisions at 62.4 GeV, I adopt $\Npart=314$ from Ref.~\cite{Braun-Munzinger:2020jbk}, while for 200 GeV, the participant number is reported as $\Npart=357\pm8$ according to Ref.~\cite{BRAHMS:2003wwg}. At 130 GeV, the $\bar{p}/p$ ratio is measured by BRAHMS for centrality bins such as 0-10\%, 10-20\%, and 20-40\% \cite{BRAHMS:2001kqb}. Additionally, around mid-rapidity, STAR reports a few data points for proton and anti-proton yields across various centralities, including 0-6\% with $\Npart=345$ \cite{STAR:2003ryp}. In this context, I estimate the net proton yields for central Au+Au collisions using the mid-rapidity $p$ from STAR multiplied by $(1-\bar{p}/p)$ from BRAHMS. To complete the procedure, estimating the net baryon distribution from the net proton distribution (or vice versa) is essential. At collision energies of 8.8, 17.3, and 62.4 GeV, factors $N^{B-\bar{B}}/N^{p-\bar{p}}$ with values of 2.71, 2.34, and 2.12, respectively, as provided in Ref.~\cite{Braun-Munzinger:2020jbk}, are applied. Additionally, at 200 GeV, a factor of $2.03\pm0.08$ is reported by BRAHMS \cite{BRAHMS:2003wwg}. For the 130 GeV collision energy, an estimated value of 2.08 is utilized based on an educated assessment. 

Utilizing available data and the methodology described above to estimate the net baryon density per participant pair from the projectile, I present the resulting distribution in Fig.~\ref{fig:baryon_scaling}. Notably, the data at 130 GeV align with the universal curve, whereas the 8.8 GeV data, not displayed in the plot, do not exhibit this behavior.\footnote{%
Note that in Ref.~\cite{STAR:2005lqw}, an alternative approach was employed to investigate the limiting fragmentation of the net proton distribution, and no universal scaling behavior was observed.
}
Consistent with Ref.~\cite{Braun-Munzinger:2020jbk}, I employ a fitting function combining a Gaussian function $a \exp[-(x - \mu)^2 / (2\sigma^2)]$ with a polynomial $c_1\,x^2+c_2\,x+c_3$. This method allows the curve fitting to capture both the peak shape (Gaussian) and additional variations (polynomial) present in the dataset. The best fit parameters are determined as $a=0.32$, $\mu=-1.65$, $\sigma=0.75$, and $c_1=-0.01$, $c_2=-0.05$, $c_3=0.05$. The fitted curve is displayed in Fig.~\ref{fig:baryon_scaling}. Upon adding back the contribution from the target, parametrized by $f_\mathrm{targ}(x)$, to the fitted curve, the resulting distribution yields the net baryon density. Using the factor $N^{B-\bar{B}}/N^{p-\bar{p}}$, the net proton distribution is estimated accordingly. The reconstructed distributions for central Au+Au collisions at 17.3, 62.4, 130, and 200 GeV are denoted by dashed lines in Fig.~\ref{fig:net_proton_reconst}. Notably, these reconstructed distributions exhibit a remarkable agreement with the measurements. Particularly noteworthy is their capacity to provide distributions with full rapidity coverage, even for the beam energies not available through experiments. 

While the efficiency of these reconstructed distributions is evident through comparison with available data, and the underlying curves are derived from well-established physics, it is crucial to handle their usage with care, particularly in Bayesian inference of parameters. This caution is necessary because these distributions are obtained without incorporating uncertainties, and extrapolating beyond the measurements would introduce non-trivial prior knowledge into the inference process. Future improvements to the reconstruction method should address uncertainties in experimental data used for fitting and consider variations in fitting functions, thereby incorporating these uncertainties into the reconstructed distributions themselves. While these improvements may be pursued in subsequent studies, the current reconstructed distributions can nevertheless provide invaluable guidance for constraining (3+1)-dimensional models, as models constrained in such a way are undoubtedly preferable to those left largely unconstrained by relying solely on midrapidity measurements (see Sec.~\ref{sec:reconst_27}). 

Finally, I note that when the theoretical model is not yet sophisticated, it is premature to constrain model parameters using Bayesian inference, even with rapidity-dependent observables. Therefore, it is essential to refine the model description before performing Bayesian calibration. This study will enhance the understanding of longitudinal dynamics in heavy-ion collisions and, consequently, facilitate improving dynamical modeling at beam energy scan energies.

\section{Results and discussion}\label{sec:results}

\subsection{Constructing unmeasured rapidity distributions}\label{sec:reconst_27}

\begin{figure}[!t]
    \centering
    \includegraphics[width= 0.8\linewidth]{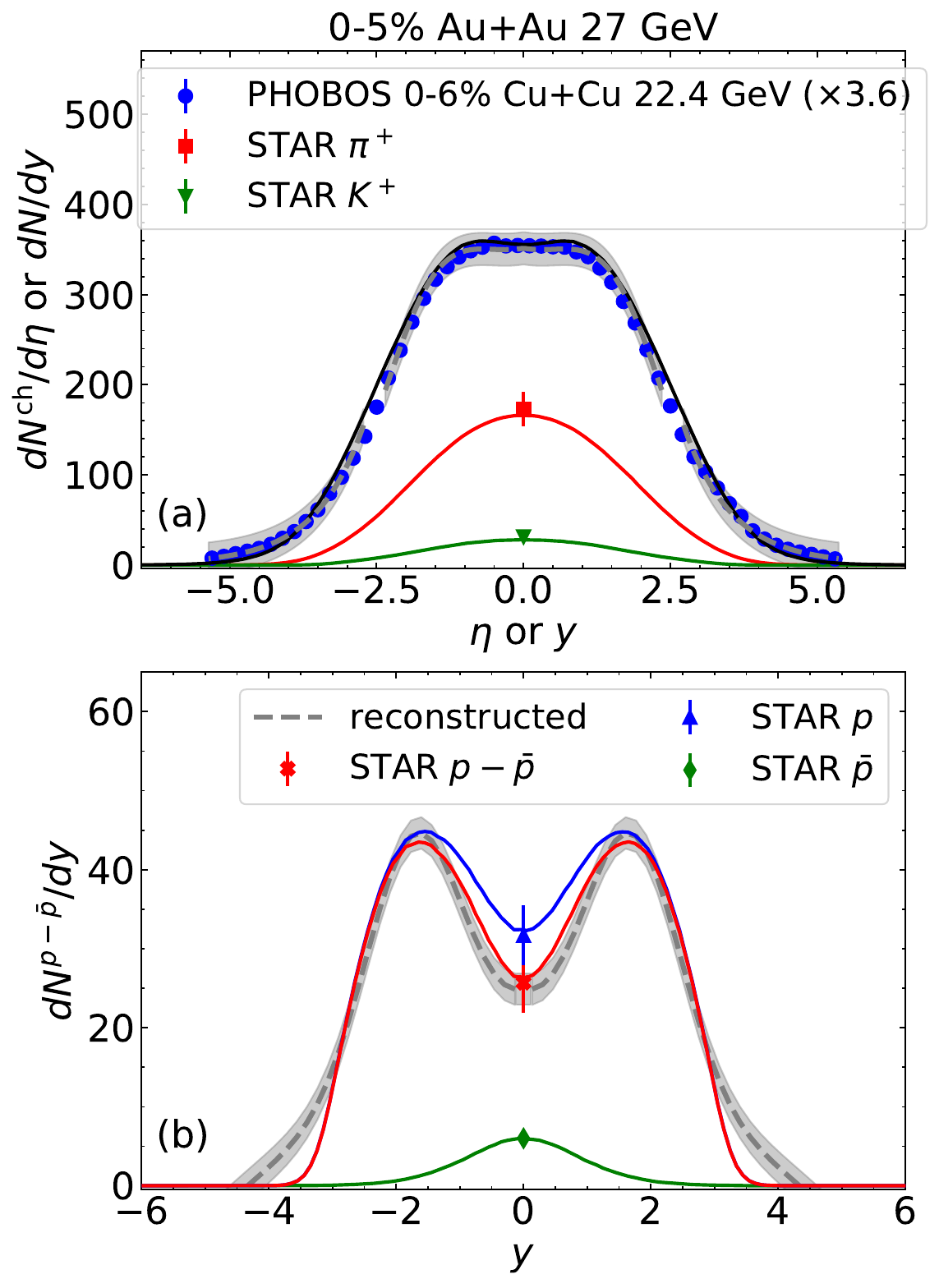}
    \caption{%
        The charged particle multiplicity in pseudo-rapidity $\eta$ and identified particle yields in rapidity $y$ for 0-5\% Au+Au collisions at 27 GeV. Experimental measurements are indicated by markers, while model calculations are illustrated by solid curves. The dashed curves with accompanying gray bands represent the reconstructed distributions for (a) charged particle multiplicity and (b) net proton yields. These distributions are derived from fitted curves as demonstrated in Figs.~\ref{fig:charged_reconst} and \ref{fig:net_proton_reconst}. The gray bands are manually added to approximate experimental uncertainties. Additionally, blue dot markers illustrate the charged particle multiplicity measured for 0-6\% Cu+Cu collisions at 22.4 GeV, rescaled by a factor estimated from participant ratios; see the text for more details.
    }
    \label{fig:reconstruct_27}
\end{figure}

As a practical application of the methodology established in Sec.~\ref{sec:rap_reconst}, I employ it to construct rapidity distributions for central Au+Au collisions at 27 GeV, where no direct rapidity measurements are available. By inserting the participant number $N_\mathrm{part}=343$ \cite{Braun-Munzinger:2020jbk} into Eq.~\eqref{eq:mid_charge}, the estimated midrapidity charged particle multiplicity is calculated as $(dN^\mathrm{ch}/d\eta)|_{\eta=0}=355$. Utilizing the beam rapidity $\yb=3.36$, I apply the procedure detailed in Sec.~\ref{sec:rap_reconst} to obtain $dN^\mathrm{ch}/d\eta$. This reconstruction is depicted in Fig.~\ref{fig:reconstruct_27}(a) as a dashed line with an accompanying band. As an additional validation, I plot the charged particle multiplicity measured for 0-6\% Cu+Cu collisions at 22.4 GeV, scaled by a participant ratio, $N_\mathrm{part}^\mathrm{Au+Au}/N_\mathrm{part}^\mathrm{Cu+Cu}$, to infer the measurements for Au+Au collisions at 27 GeV.\footnote{%
The validity of this approximation is demonstrated in Appendix~\ref{app:aucu}.
}
Notably, the rescaled multiplicity observed in Cu+Cu collisions at 22.4 GeV aligns remarkably well with the reconstruction at 27 GeV, which is anticipated due to the similar $\yb$ between the two beam energies.

Moreover, I can derive the net proton distribution employing the established methodology. For the net proton distribution, at 27 GeV, given the availability of the net proton yield at midrapidity, I can scale the constructed net baryon distribution to ensure that the resulting distribution accurately reproduces the measured net proton yield at midrapidity. The reconstructed net proton distribution is represented as a dashed line with a band in Fig.~\ref{fig:reconstruct_27}(b). Leveraging the obtained $dN^\mathrm{ch}/d\eta$ and $dN^{p-\bar p}/dy$, I calibrate the dynamical model and the results are shown as solid lines in Fig.~\ref{fig:reconstruct_27}. Notably, when reproducing the reconstructed $dN^\mathrm{ch}/d\eta$, the model inherently reproduces the measured pion and kaon yields at midrapidity from STAR as well. Figure~\ref{fig:reconstruct_27} exemplifies the efficacy of the established method in constraining longitudinal dynamics, particularly when dealing with limited rapidity data.

\subsection{Nuclear remnants and string junction}\label{sec:fragment}

As discussed earlier, the measurements of rapidity-dependent quantities such as $dN^\mathrm{ch}/d\eta$ and $dN^{p-\bar p}/dy$ serve as valuable constraints for understanding bulk dynamics, particularly in the beam direction, across the beam energy scan collisions. Leveraging these rapidity datasets alongside a calibrated dynamical model, I can estimate the total energy and baryon number deposited within the entire collision fireball, which aids the understanding of initial energy deposition and baryon stopping mechanisms. Specifically examining collisions at 19.6 GeV, where rapidity measurements are accessible across various centralities, the objective in this section is to offer theoretical insights into nuclear remnants and string junctions.

\begin{figure}[!tb]
    \centering
    \includegraphics[width= \linewidth]{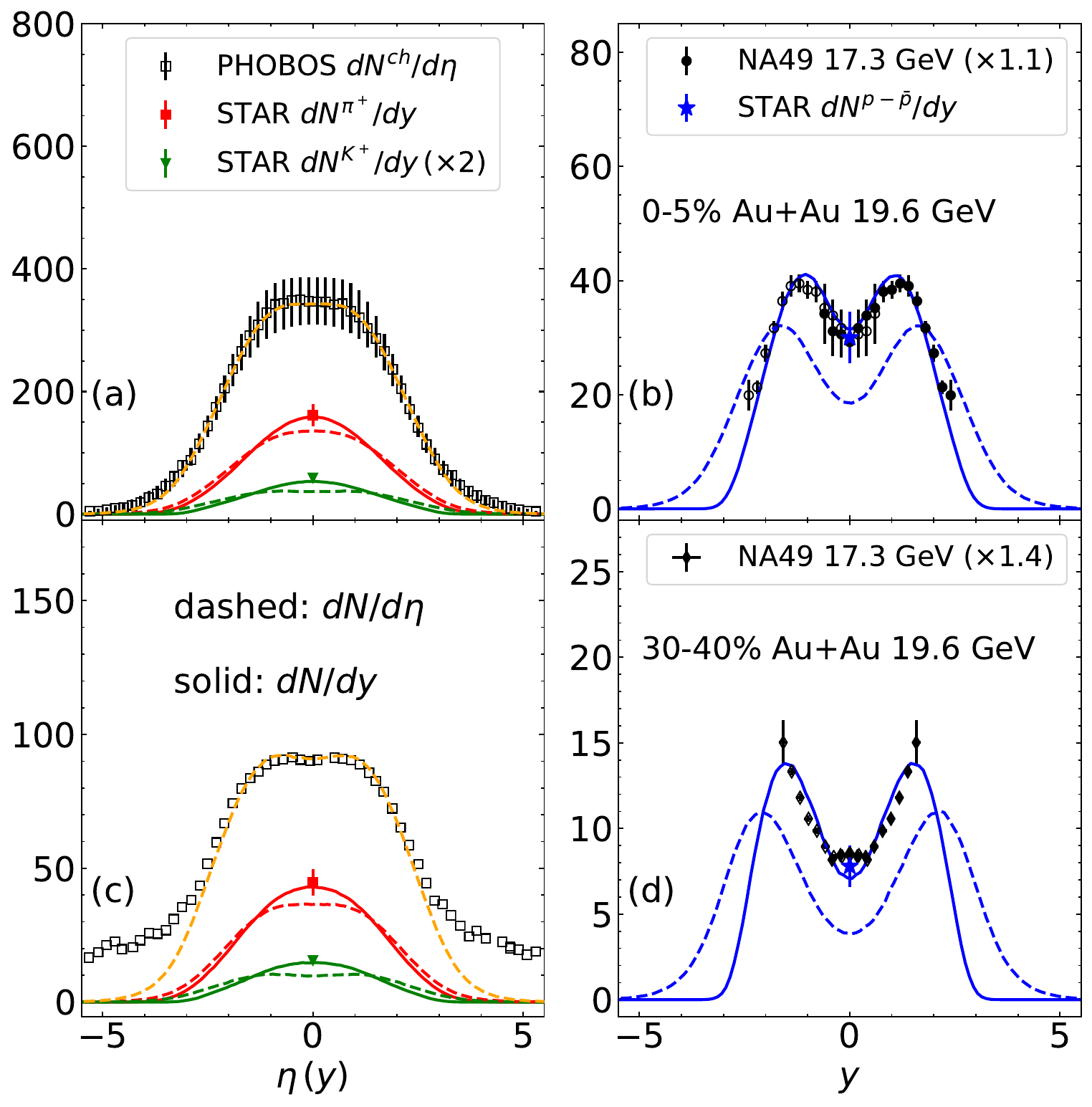}
    \caption{%
        The charged particle multiplicity in pseudo-rapidity $\eta$ and identified particle yields in rapidity $y$ for (a,\,b) 0-5\% and (c,\,d) 30-40\% Au+Au collisions at 19.6 GeV. Experimental measurements are indicated by markers, while model calculations are illustrated by curves. The solid lines represent rapidity density $dN/dy$, while the dashed lines denote pseudo-rapidity density $dN/d\eta$. The net proton distributions obtained from Pb+Pb collisions at 17.3 GeV, as measured by NA49, are adjusted by a constant factor to align with the midrapidity net proton yields from Au+Au collisions at 19.6 GeV measured by STAR. A factor is applied to the kaon yields only to enhance the clarity of the illustration. Experimental uncertainties of $dN^\mathrm{ch}/d\eta$ are not plotted in (c).
    }
    \label{fig:rapidity_dist_19}
\end{figure}

Figure~\ref{fig:rapidity_dist_19} demonstrates my approach: I utilize the charged particle multiplicity within 0-5\% and 30-40\% centralities measured by PHOBOS, offering extensive rapidity coverage, to calibrate entropy and energy evolution. To cross-validate, I employ the measured pion and kaon yields at midrapidity from STAR. Furthermore, I adjust the net proton distributions obtained from Pb+Pb collisions at 17.3 GeV measured by NA49 with a constant factor. This adjustment aligns these distributions at midrapidity with the net proton yields from Au+Au collisions at 19.6 GeV, measured by STAR.\footnote{%
Indeed, I had the option to employ the procedures outlined in Sec.~\ref{sec:rap_reconst} to reconstruct the net proton distribution in rapidity at 19.6 GeV. This reconstructed distribution would align with the scaled measurements at 17.3 GeV, considering the very similar $\yb$ between these two beam energies.
}
Subsequently, these resulting distributions aid in calibrating the evolution of baryon density.

For 0-5\% Au+Au collisions, to obtain the distributions in Fig.~\ref{fig:rapidity_dist_19}(b) in the model calculation, I have initialized the system with a participant number calculated as $\Npart=335$. This number aligns well with the $\Npart=337\pm12$ reported for 0-6\% centrality by PHOBOS \cite{Back:2002wb}. Interestingly, the net baryon number during the hydrodynamic stage is calculated as $N_B=333$, and thus $N_B\approx\Npart$. Similarly, for collisions within 30-40\% centrality, I observe $N_B\approx\Npart=111$, consistent with the participant number provided by PHOBOS \cite{Back:2002wb}. Admittedly, the rapidity distribution measurement in this instance, as depicted in Fig.~\ref{fig:rapidity_dist_19}(d), remains incomplete. The approximate equality between $N_B$ and $\Npart$, suggests that the total baryon number carried by the participants becomes deposited into the collision fireball. The assumption made in Sec.~\ref{sec:rap_reconst}, considering $N_B$ to be equivalent to $\Npart$, has found further support through the current model-to-data comparison.

Comparing the incoming energy carried by participants, $E_\mathrm{part}=\Npart\snn/2$, and the total energy during hydrodynamic evolution, $E_\mathrm{hydro}=\int d^3\Sigma_\mu T^{\mu t}(\bm{x}_\perp,\,\eta_s)$ at a constant proper time, offers valuable insights into the energy deposition during the initial collision stage. For instance, in Ref.~\cite{Shen:2020jwv}, it is assumed that these two total energies are identical, a constraint not explicitly embedded in the model. Nonetheless, I find that these two total energies exhibit agreement within a discrepancy of less than 15\% for 0-5\% centrality. For 30-40\% centrality, tuning the initial condition to equalize $E_\mathrm{hydro}$ with $E_\mathrm{part}$ leads to $dN^\mathrm{ch}/d\eta$ depicted by the dashed line in Fig.~\ref{fig:rapidity_dist_19}(c). Apparently, this results in a considerable underestimation of the measured charged particle multiplicity in the fragmentation region by the model calculation.
Yet, reproducing the extensive tails in the distribution necessitates a higher energy contribution than from incoming participants, particularly since particles at larger pseudo-rapidities carry substantial longitudinal momenta. This observation suggests that these tails likely stem from charged particles emitted by excited nuclear remnants, rather than originating from the participant collision fireball. The presence of remnants at large pseudorapidity has indeed been validated by PHOBOS measurements~\cite{PHOBOS:2015aeh}. Here, I offer a theoretical approach to distinguish and identify nuclear remnants. 

Identifying nuclear remnants holds significant importance in comprehending the directed flows of charged particles in pseudo-rapidity, $v^\mathrm{ch}_1(\eta)$. Notably, if there is a substantial contribution to charged particles from nuclear remnants at forward (backward) rapidities, $v^\mathrm{ch}_1(\eta)$ could exhibit notably positive (negative) values. Merely interpreting $v^\mathrm{ch}_1(\eta)$ at forward and backward rapidities through a pure hydrodynamic model describing the collision fireball evolution would be insufficient. Furthermore, to interpret the sign of $v^\mathrm{ch}_1(\eta)$ by considering the signs of $v_1(y)$ for identified particles measured in rapidity ($y$), it becomes crucial to understand the contributions of identified particles at various pseudo-rapidities ($\eta$). Figure~\ref{fig:rapidity_dist_19} also depicts the pseudo-rapidity distributions of identified particles using dashed lines, notably differing from those in rapidity,\footnote{%
This contradicts the arguments in Ref.~\cite{PHOBOS:2010eyu} to approximate $y$ by $\eta$ and $dN/dy$ by $dN/d\eta$ in the fragmentation region.
}
particularly observed for protons.
Therefore, it is crucial to avoid interpreting $v_1(y)$ for identified particles as equivalent to $v_1(\eta)$ when assessing the sign and value of $v^\mathrm{ch}_1(\eta)$ for charged particles.

Conducting systematic exploration to understand the relationship between the energy carried by  incoming participants, $E_\mathrm{part}$, and the total energy during hydrodynamic evolution, $E_\mathrm{hydro}$, holds potential for comprehending the initial energy deposition and baryon stopping mechanisms. In the traditional baryon deceleration picture, the energy deposited into the fireball is typically attributed to the decelerated incoming nucleons and their energy loss. This picture would yield a strong correlation between the net proton distribution in rapidity (governed by baryon deceleration and energy loss) and the distribution of charged particle multiplicity (reflecting the energy deposition into the fireball). However, if a significant fraction of the net baryon number originates from the breaking of string junctions  \cite{Kharzeev:1996sq, Sjostrand:2002ip}, the energy deposition could notably deviate from the baryon deceleration concept. Such deviation might disrupt the expected correlation between the net proton distribution and the distribution of charged particle multiplicity. The equivalence between $E_\mathrm{part}$ and $E_\mathrm{hydro}$ suggests a lack of strong correlation between the deposited energy and the deceleration (or energy loss) experienced by incoming participants. This observation raises the question of whether there exists a significant contribution to baryon production resulting from string junction breaking, necessitating more systematic exploration.

Additional insights into the string junction scenario could potentially be obtained from the scaling properties in the net baryon distribution illustrated in Fig.~\ref{fig:baryon_scaling}. The universal curve depicted in the figure was derived by parametrizing the target contribution as an average of two exponential functions scaled by a constant factor, namely $C[\exp(-x)+\exp(-x/2)]/2$ in Eq.~\eqref{eq:targ}. Notably, these two functions correspond to two distinct baryon distributions: $\exp(-x)$ from Ref.~\cite{Busza:1983rj}, and $\exp(-x/2)$, motivated by a gluon junction picture \cite{Kopeliovich:1988qm}. It would be overly simplistic to directly associate the observed universal curve with the presence of the string junction, considering that multiple dynamic effects, in particular baryon transport, can significantly influence the final net proton distribution.\footnote{%
Note, however, that Ref.~\cite{STAR:2005lqw} argued that the absence of such a universal curve would potentially suggest the existence of the string junction, which contradicts the argument presented here.
}
However, this observation could potentially serve as motivation for a systematic exploration of various baryon stopping scenarios using rapidity data of net protons in a similar way. This systematic exploration may shed light on the additional rapidity-independent component within the net baryon distribution introduced in Ref.~\cite{Du:2022yok} that could be attributed to the string junction. This line of exploration may offer valuable insights into the underlying mechanisms governing baryon transport and stopping in heavy-ion collisions.

\subsection{Probing the Equation of State}\label{sec:eos}

The nuclear matter produced in heavy-ion collisions conserves multiple charges, such as baryon number ($B$), electric charge ($Q$), and strangeness ($S$), which are correlated in their evolution. Describing this interplay involves understanding the EoS, which defines the relationship between thermodynamic parameters in a four-dimensional space, $p(T,\,\mu_B,\,\mu_Q,\,\mu_S)$, linking pressure ($p$) with temperature ($T$) and three chemical potentials associated with $BQS$-charges. Currently, integrating the EoS into hydrodynamic models often necessitates simplifications by projecting the four-dimensional EoS onto a two-dimensional space. This involves imposing constraints that interrelate the strangeness and electric charge with baryon number. For instance, EoS like NEOS-B assumes vanishing strangeness and electric charge chemical potentials ($\mu_S {\,=\,} \mu_Q {\,=\,} 0$), while NEOS-BQS implies strangeness neutrality ($n_S{\,=\,}0$) and maintains a fixed electric charge-to-baryon ratio ($n_Q{\,=\,}0.4\,n_B$)\footnote{
The electric charge-to-baryon ratio value of 0.4, utilized in NEOS-BQS, is specifically designed for collisions involving heavy nuclei like Au or Pb. Extending NEOS-BQS to small colliding systems is valuable for future investigation, albeit beyond the scope of the present study.
} 
\cite{Monnai:2019hkn,Monnai:2021kgu}. These constraints allows for the focus solely on baryon charge evolution while deriving the evolution of electric charge and strangeness via the embedded constraints within the EoS \cite{Monnai:2019hkn,Du:2022yok}.

\begin{figure}[!tb]
    \centering
    \includegraphics[width= 0.86\linewidth]{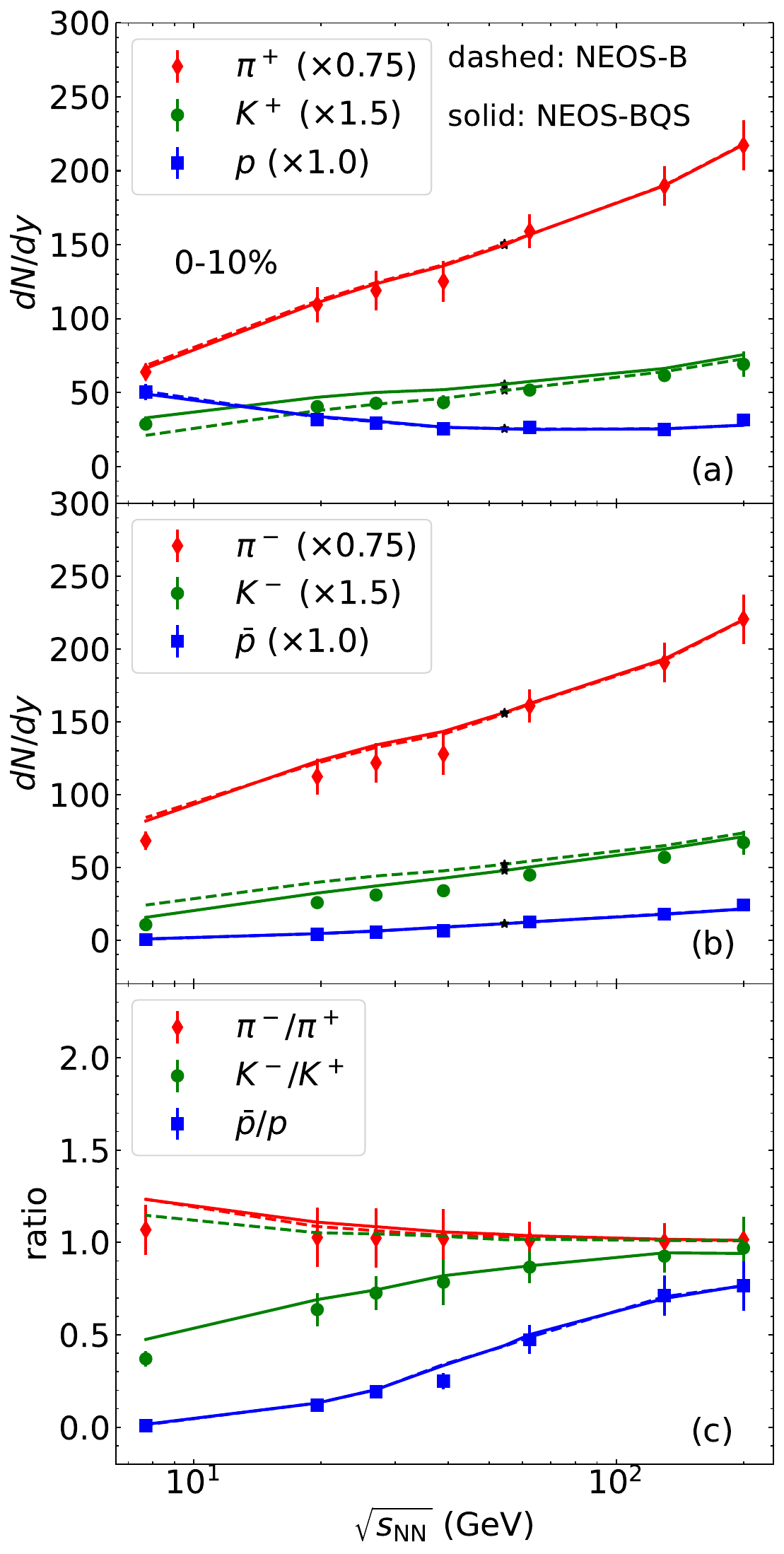}
    \caption{Yields of identified particles at midrapidity ($|y|<0.1$), with (a) positive and (b) negative charges, alongside (c) the corresponding ratios for 0-10\% Au+Au collisions across beam energies from 7.7 to 200 GeV. STAR measurements are represented by markers while model calculations are depicted by lines. The dashed lines correspond to results using NEOS-B, while the solid lines represent those with NEOS-BQS. Various factors are applied to the identified particle yields to enhance the clarity of the illustration.}
    \label{fig:BES_yields}
\end{figure}

In Ref.~\cite{Monnai:2019hkn}, it was demonstrated that introducing a non-zero $\mu_S$, in contrast to assuming $\mu_S {\,=\,}0$, leads to better agreement between theoretical calculations and experimental measurements around midrapidity for most particles with strangeness in Pb+Pb collisions at 17.3 GeV. Extending this comparison to various beam energies at BES, as depicted in Fig.~\ref{fig:BES_yields}, reveals distinct differences in identified particle yields around midrapidity obtained from NEOS-B (dashed lines) and NEOS-BQS (solid lines). As illustrated in Fig.~\ref{fig:BES_yields}(a,b), the disparities primarily manifest in the yields of $K^{\pm}$, which carry strangeness, while the yields of $\pi^{\pm}$, $p$, and $\bar p$ are minimally affected. This aligns with the observations from Ref.~\cite{Monnai:2019hkn} for the 17.3 GeV collisions. A noticeable enhancement in the agreement between theoretical calculations and experimental measurements emerges in the particle yield ratio $K^-/K^+$ as a function of beam energy when utilizing NEOS-BQS, as illustrated in Fig.~\ref{fig:BES_yields}(c). With NEOS-B, the yield ratio $K^-/K^+$ increases as the beam energy decreases, contrary to the trend observed in measurements. Remarkably, using NEOS-BQS reproduces this trend accurately, aligning more closely with experimental measurements.

The reason for the notable enhancement in the $K^-/K^+$ ratio when employing NEOS-BQS has been extensively discussed in Ref.~\cite{Shen:2022oyg}, where a dynamic initialization method was employed for the hydrodynamic evolution. For the convenience of readers, I summarize it here. The particlization process, where the Cooper-Frye prescription \cite{Cooper:1974mv} is utilized, assumes a grand canonical ensemble in this study. Due to the distinct quantum numbers carried by various hadron species (such as $K^+$ with hadronic chemical potentials $\mu_Q+\mu_S$ and $K^-$ exhibiting the opposite), the yield ratio $K^-/K^+$ is proportional to $\exp[-2(\mu_Q+\mu_S)]$. In the NEOS-BQS, strangeness neutrality ($n_S{\,=\,}0$) establishes a correlation between $\mu_S$ and $\mu_Q$, specifically $\mu_S\approx\mu_B/3$, while maintaining a fixed electric charge-to-baryon ratio requiring $n_Q{\,=\,}0.4\,n_B$. Consequently, as the beam energy decreases and both $n_B$ and $\mu_B$ increase, the corresponding increase in $\mu_Q$ and $\mu_S$ leads to the suppression of $K^-/K^+$. It is important to note that a definitive conclusion regarding the suppression or enhancement of $\pi^\pm$, $p$, and $\bar p$ yields by NEOS-BQS cannot be drawn, as illustrated in Fig.~\ref{fig:BES_yields}(a,b).

\begin{figure}[!tb]
    \centering
    \includegraphics[width= 0.86\linewidth]{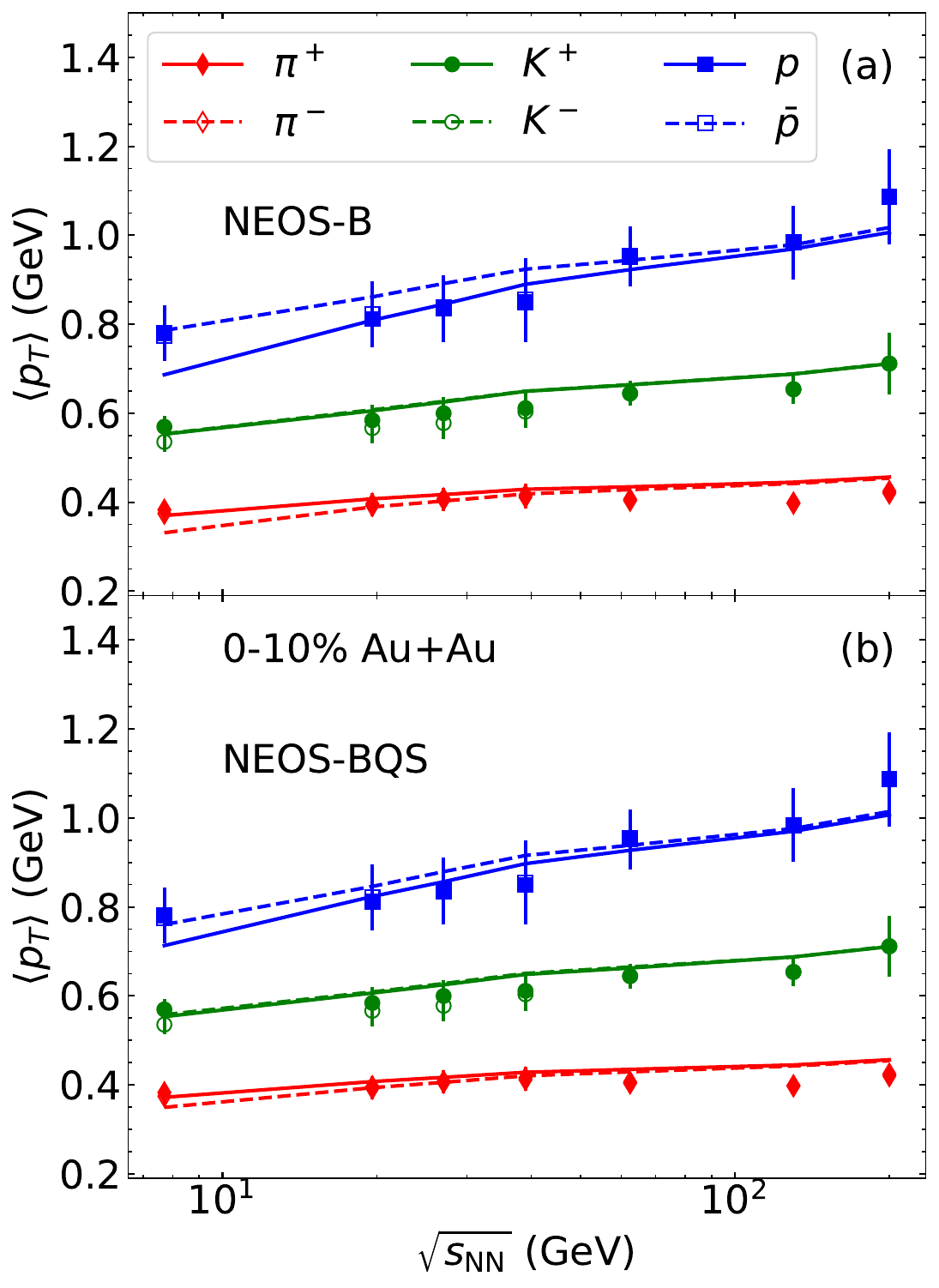}
    \caption{Mean $p_T$ of identified particles at midrapidity ($|y|<0.1$), with positive (solid markers or solid lines) and negative (hollow markers or dashed lines) electric charges for 0-10\% Au+Au collisions across beam energies from 7.7 to 200 GeV. STAR measurements are represented by markers while model calculations are depicted by lines. The upper panel corresponds to results using NEOS-B, while the lower panel represents those with NEOS-BQS.}
    \label{fig:BES_pT}
\end{figure}

In Fig.~\ref{fig:BES_pT}, the model results for the mean transverse momentum of identified hadrons at midrapidity are presented across various beam energies using NEOS-B (upper panel) and NEOS-BQS (lower panel). Both positively and negatively charged hadrons are shown in each panel. Notably, in both panels, the calculated mean $p_T$ of anti-protons consistently exceeds that of protons across the beam energies,\footnote{%
The higher mean $p_T$ of anti-protons compared to protons, as explained in Ref.~\cite{Denicol:2018wdp}, results from three factors. First, the value of $\mu_B/T$ decreases during the earliest hydrodynamic evolution when radial flow begins to develop. This anticorrelation between $\mu_B/T$ and flow in the early stages leads to a relatively higher production of protons when the radial flow is minimal. Second, baryon diffusion plays a role, driven by the $\mu_B/T$ gradient in the transverse plane, which tends to diffuse net-baryon charge into the central region where radial flow is weaker. Third, the baryon diffusion $\delta f$ corrections to the baryon spectra contribute significantly to this difference.
} 
with a more pronounced difference at lower beam energies associated with higher chemical potentials, consistent with findings from Ref.~\cite{Shen:2020jwv}. An intriguing observation is the reduction in the disparity between the mean $p_T$ of protons and anti-protons when employing NEOS-BQS. This reduction results in a closer alignment with experimental measurements, wherein the mean $p_T$ for both protons and anti-protons demonstrate agreement within the uncertainties. Additionally, a striking consistency emerges between the calculated mean $p_T$ of $K^+$ and $K^-$, showing minimal deviation despite the choice of different equations of state. This contrasts with the variations observed in the yields of $K^+$ and $K^-$ depicted in Fig.~\ref{fig:BES_yields}. Overall, the utilization of NEOS-BQS not only enhances the yields but also refines the agreement with measurements in the mean $p_T$ values across various identified species around midrapidity.

Drawing from the discussion above, it might seem apparent that NEOS-BQS stands as the superior choice for simulating heavy-ion collisions at finite chemical potentials. However, it is crucial to acknowledge that the two embedded constraints are based on two considerations: the vanishing strangeness density in nucleons and the fixed $n_Q/n_B=Z/A$ ratio, where $Z/A\approx0.4$ for nuclei like Au or Pb, both applicable to averaged quantities across the colliding nuclei. Throughout the hydrodynamic evolution, these multiple charges evolve in correlation, yet these two constraints might not hold locally. For instance, while the total strangeness of the entire system remains zero, the local strangeness could vary from point to point \cite{Fotakis:2022usk,Carzon:2019qja,Carzon:2023zfp}. In the NEOS-BQS, these constraints are enforced on each fluid cell, potentially leading to overly stringent limitations.

Rapidity-dependent measurements provide a clearer illustration of this issue. In a recent study by Ref.~\cite{Du:2022yok}, the implementation of NEOS-BQS was observed to disrupt the agreement between theoretical predictions and experimental measurements, particularly in the directed flows in rapidity, $v_1(y)$, for identified hadrons with strangeness (e.g., $\Lambda$ and $K^+$ carrying opposite $\mu_S$). According to the experimental measurements at 7.7 GeV, the slope of $v_1(y)$ around midrapidity showed opposite signs for $\Lambda$ and $K^+$. The enforced strangeness neutrality within NEOS-BQS tends to couple their $v_1(y)$, leading to a similar rapidity dependence, contradicting the observed measurements. This study underscores the necessity of hydrodynamically evolving multiple charges by employing the complete four-dimensional EoS, $p(T,\,\mu_B,\,\mu_Q,\,\mu_S)$, when investigating some particular rapidity-dependent observables.

\begin{figure}[!tb]
    \centering
    \includegraphics[width= \linewidth]{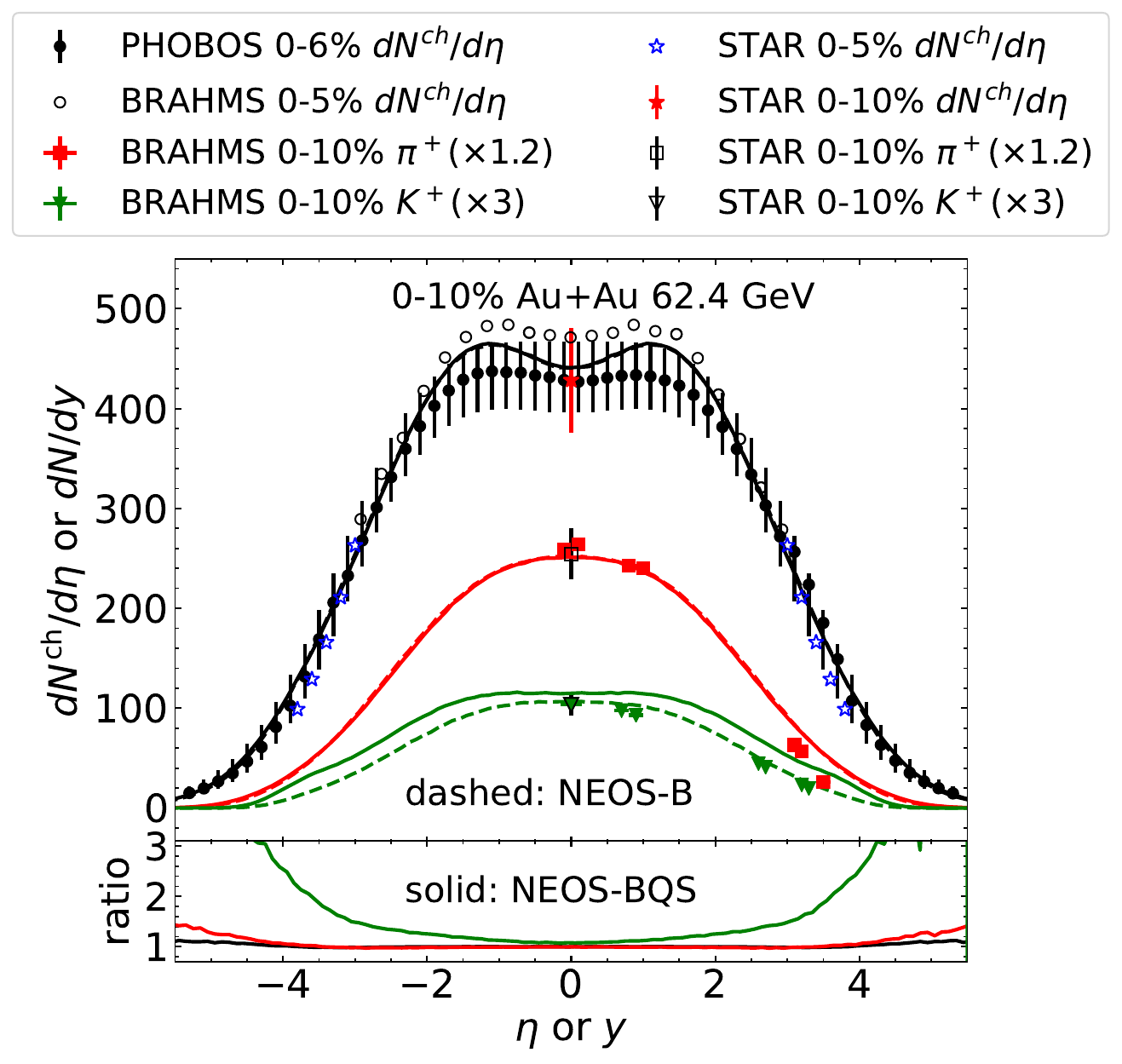}
    \caption{(Upper) The charged particle multiplicity in pseudo-rapidity $\eta$ and identified particle yields in rapidity $y$ for 0-10\% Au+Au collisions at 62.4 GeV. Experimental measurements are represented by markers while model calculations are depicted by lines. The  charged particle measurements from various collaborations are plotted as a reference for readers. The dashed lines correspond to NEOS-B, while the solid lines represent NEOS-BQS. Various factors are applied to the identified yields to enhance the clarity of the illustration. (Lower) The ratio between distributions using NEOS-BQS and NEOS-B, where the latter serves as the denominator in the ratio.}
    \label{fig:rap_62_BQS}
\end{figure}

Investigating rapidity-dependent yields offers an extra method to explore the EoS at finite chemical potentials. For instance, I analyze central Au+Au collisions at 62.4 GeV, leveraging available rapidity-dependent data for identified particle yields. In Fig.~\ref{fig:rap_62_BQS}, observing the lower panel depicting the ratios between distributions employing NEOS-BQS and NEOS-B, I note a distinct enhancement in $K^+$ yields across rapidity due to NEOS-BQS. However, definitive conclusions regarding the suppression or enhancement of $\pi^+$ yields across rapidity using NEOS-BQS remain elusive. This echoes the findings for midrapidity yields across various beam energies shown in Fig.~\ref{fig:BES_yields}. The $dN/dy$ profile of $K^+$ becomes irregular with NEOS-BQS, disrupting its smooth distribution with NEOS-B.\footnote{%
This effect becomes more prominent at lower beam energies, notably at 7.7 GeV, a result I verified but did not explicitly include in the displayed figures, as no rapidity data are available.
} 
The appearance of double-humped $\mu_B$ in rapidity, aimed at reproducing the measured net proton distribution, alongside the correlation between $\mu_S$ and $\mu_B$, contributes to the increased irregularity in $dN/dy$ profile of $K^+$. This observation underscores the potential significance of employing rapidity-dependent yields of identified hadrons in probing the EoS at finite chemical potentials.

\subsection{Thermodynamic properties at freeze-out}\label{sec:freezeout}

Demonstrating the multistage framework's ability to reproduce various experimental measurements, particularly the identified particle yields and mean $p_T$ around midrapidity across beam energies, I further investigate thermodynamic properties like the freeze-out temperature and baryon chemical potential on the freeze-out hypersurface at the particlization process. This exploration offers valuable insights into comprehending the extracted freeze-out parameters from thermal models using identified particle yields \cite{Braun-Munzinger:2003pwq,Tawfik:2014eba,STAR:2017sal,Andronic:2017pug}.

It has been demonstrated that thermodynamic properties, especially the baryon chemical potential, undergo significant variations across spacetime rapidity ($\eta_s$), particularly at lower beam energies \cite{Du:2023gnv}. Thermal smearing results in a rapidity spread approximately of the order $\sqrt{T/\langle m_T\rangle}$, causing particle yields at a specific rapidity to originate from nuclear matter spanning various spacetime rapidities \cite{Schnedermann:1993ws,Begun:2018efg}. Consequently, extracted freeze-out parameters might represent averaged values across a broad spacetime rapidity range, even if the particle yields used for extraction are confined to a very narrow window around midrapidity, such as $|y|<0.1$. In this section, I illustrate this concept further by comparing the freeze-out temperatures ($T$) and baryon chemical potentials ($\mu_B$) across various beam energies between the values on the hydrodynamic hypersurface and those derived from the thermal models.

\begin{figure}[!tb]
    \centering
    \includegraphics[width= 0.85\linewidth]{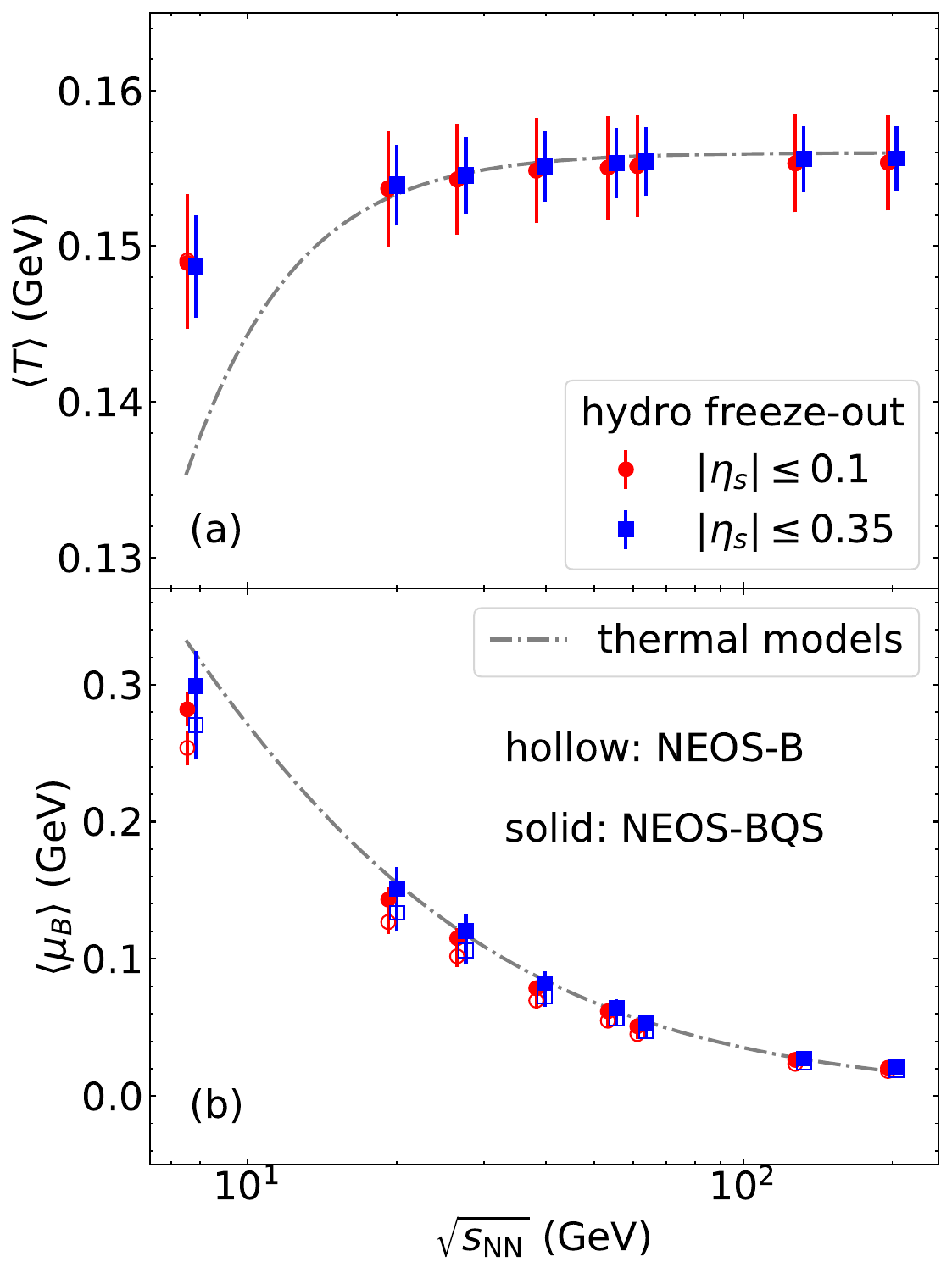}
    \caption{Freeze-out (a) temperature $T$ and (b) baryon chemical potential $\mu_B$ around midrapidity on the hydrodynamic freeze-out hypersurface characterized by a constant energy density $e_\mathrm{fo}=0.35\,$GeV/fm$^3$. Markers with error bars depict the mean and standard deviation values for the energy-density weighted $T$ and baryon-density weighted $\mu_B$. Solid markers represent results using NEOS-BQS, while hollow markers indicate results using NEOS-B. Two spacetime rapidity windows, centered around midrapidity, are imposed on the fluid cells: circle markers denote $|\eta_s|\leq0.1$, and square markers depict $|\eta_s|\leq0.35$. The markers for the two windows have been slightly horizontally shifted to enhance the clarity. The dot-dashed line represents parametrizations for thermal model results from Ref.~\cite{Andronic:2017pug}, with adjusted parameters. 
    }
    \label{fig:BES_freezeout}
\end{figure}

To determine the freeze-out temperatures and baryon chemical potentials on the hydrodynamic hypersurface ($e_\mathrm{fo}=0.35\,$GeV/fm$^3$), I compute the mean and standard deviation of energy-density weighted temperatures and baryon-density weighted chemical potentials for fluid cells within a spacetime rapidity window, depicted by markers with error bars in Fig.~\ref{fig:BES_freezeout}.\footnote{%
It is worth noting that Ref.~\cite{Du:2023gnv} demonstrates the error bars on the markers using distribution percentiles, highlighting their non-Gaussian nature.
}
The weighting considers each fluid cell's energy ($e$) and baryon ($n_B$) densities, acknowledging their varying contributions to final particle yields and net-baryon numbers.\footnote{%
Variances in the choice of weight among studies can yield slightly varied results. For instance, certain studies use $\gamma e$ as a weight, where $\gamma$ denotes the Lorentz factor \cite{Churchill:2023vpt,Churchill:2023zkk}.
}
From the perspective of thermal models, extracted $T$ and $\mu_B$ values from final yields should reflect fluid cells with higher energy and baryon densities. To demonstrate the impact of changing thermodynamic properties along $\eta_s$, I also compute the results for two $\eta_s$ windows centered around midrapidity, specifically $|\eta_s|\leq0.1$ (dot markers) and 0.35 (square markers).

Figure~\ref{fig:BES_freezeout} demonstrates a decrease in $T$ alongside an increase in $\mu_B$ as the beam energy decreases, reflecting the characteristics of both the phase transition line (with lower $T$ at higher $\mu_B$) and the hydrodynamic freeze-out line characterized by a constant energy density. In the narrow $\eta_s$ window ($|\eta_s|\leq0.1$), the error bars exhibit noticeable fluctuations in $T$ and $\mu_B$, representing changing properties across the transverse plane. Widening the $\eta_s$ window ($|\eta_s|\leq0.35$) results in a more pronounced increase in $\mu_B$, while its effect on $T$ remains marginal. This is because of the stronger variations of increasing $\mu_B$ towards forward and backward rapidities, contrasting with the comparatively smaller changes in decreasing $T$ \cite{Du:2023gnv}. These observations hold true for the hydrodynamic results using both NEOS-B (hollow markers) and NEOS-BQS (solid markers).

I represent the parametrizations of $T$ and $\mu_B$ from thermal models as functions of $\snn$, following the equations $T=T_0/\{1+\exp[2.60-\ln(\snn)/0.45]\}$ and $\mu_B=a/(1+0.288\snn)$, where $\snn$ is in GeV. While Ref.~\cite{Andronic:2017pug} uses parameters $T_0=0.1584\,$GeV and $a=1.3075\,$GeV, I adopt $T_0=0.156\,$GeV and $a=1.05\,$GeV, resulting in improved agreement with the results at higher beam energies, as indicated by the dot-dashed lines in Fig.~\ref{fig:BES_freezeout}. These adjustments only alter the overall scales of $T$ and $\mu_B$, preserving the functional shapes across the range of considered beam energies. The parametrizations align closely with extracted $T$ and $\mu_B$ values obtained from thermal models based on midrapidity yields for central collisions \cite{Andronic:2017pug}. Therefore, the comparison between the dot-dashed lines and the markers serves as a comparison between hydrodynamic freeze-out and thermal models.

Figure~\ref{fig:BES_freezeout} illustrates a notable discrepancy in the freeze-out chemical potentials between the hydrodynamic results using NEOS-BQS and those using NEOS-B, which makes results using NEOS-BQS align more closely with the dot-dashed line representing the thermal model results. However, there is only a marginal change in temperatures between the two hydrodynamic results. An intriguing observation is the discrepancy between hydrodynamic and thermal model results: while good agreement is observed above 19.6 GeV, the hydrodynamic results yield higher temperatures and smaller chemical potentials than the thermal models at 7.7 GeV.\footnote{%
It is important to note that the comparison between these discrepancies of $T$ and $\mu_B$ should account for the substantial difference in the $y$-axis ranges between the two plots.
}
This deviation is likely attributable to thermal smearing effects and the variation in thermodynamic properties of the collision fireball across spacetime rapidity \cite{Du:2023gnv}. 

As previously discussed, due to thermal smearing, particle yields measured within a small rapidity window around midrapidity can originate from a significantly wider spacetime rapidity range. For example, in the case of net protons, a significant contribution might emerge from nuclear matter situated away from $\eta_s=0$, characterized by larger chemical potentials \cite{Du:2023gnv}. Consequently, when employing $dN^{p-\bar p}/dy$ within $|y|<0.1$, thermal models provide a $\mu_B$ averaged over a broader $\eta_s$ window, resulting in a larger $\mu_B$ compared to that at $\eta_s=0$. Conversely, such averaging across a wider $\eta_s$ window leads to a smaller temperature compared to the $T$ at $\eta_s=0$. Expanding the $\eta_s$ window at 7.7 GeV is expected to narrow the discrepancy between the hydrodynamic results and the extraction of thermal models. This explanation clarifies the disparities seen in Fig.~\ref{fig:BES_freezeout} between the hydrodynamic and thermal model results, underscoring the significant influence of the thermal smearing effect.

\section{Summary and conclusions}\label{sec:conclusions}

Utilizing available rapidity data across a wide range of beam energies, I consolidate existing methods and propose additional methodologies to discern universal scaling properties among these data points (Sec.~\ref{sec:rap_reconst}). Using this established approach, one can systematically reconstruct full rapidity distributions for charged particle multiplicity and net proton yields, even at beam energies where measurements are absent. These reconstructed distributions play an essential role in constraining longitudinal dynamics, as demonstrated for the collisions at 27 GeV (Sec.~\ref{sec:reconst_27}).

Moreover, I discuss the potential of rapidity distributions in comprehending energy deposition and baryon stopping mechanisms, as demonstrated for the collisions at 19.6 GeV (Sec.~\ref{sec:fragment}). With distributions of net proton and charged particle multiplicity, I gain insights into identifying nuclear remnants in the fragmentation region and understanding mechanisms related to energy deposition and baryon stopping. Specifically, understanding the total energy deposited into the fireball can help differentiate between baryon stopping due to incoming nucleon deceleration and string junction breaking.

I also explore the impact of different equations of state and underscore the effects of considering nonzero $\mu_S$ (Sec.~\ref{sec:eos}). Using NEOS-BQS notably improves yield ratios for hadron species with strangeness and enhances the mean $p_T$ of charged hadrons consistently across various beam energies. However, the imposition of local strangeness neutrality within NEOS-BQS imposes constraints affecting the agreement between model predictions and experimental measurements, notably observed in $v_1(y)$ for identified hadrons with strangeness, as discussed in Ref.~\cite{Du:2022yok}. Furthermore, this study indicates that the utilization of NEOS-BQS may introduce irregularities in $dN/dy$ distributions of identified hadrons with strangeness due to their coupling with net-baryon distributions, originating from the constraints embedded in the EoS. This underscores the essential role of rapidity-dependent measurements in probing the equations of state at finite chemical potentials.

Utilizing the calibrated multistage framework, I investigate the thermodynamic properties at the hydrodynamic freeze-out hypersurface around midrapidity, emphasizing the combined effects of thermal smearing and variations in thermodynamic properties across rapidities (Sec.~\ref{sec:freezeout}). Notably, I observe that wider spacetime rapidity windows correspond to higher mean baryon chemical potentials. Upon comparison with parametrizations representing thermal model results, discrepancies emerge, particularly at lower beam energies. These discrepancies highlight the implications of thermal smearing and longitudinal thermodynamic variations. This emphasizes that the freeze-out parameters extracted from thermal models, based on midrapidity measurements, are essentially averaged properties across rapidities.

In summary, this study emphasizes the significance of rapidity-dependent measurements for model calibration at beam energy scan and revealing QCD properties at finite chemical potentials. The calibrated multistage framework investigated here can aid in interpreting experimental measurements in the second stage of the beam energy scan program.

\section*{Acknowledgments}

The author acknowledges fruitful discussions with Charles Gale, Ulrich Heinz, Sangyong Jeon, and Chun Shen. The author extends gratitude to Charles Gale and Sangyong Jeon for their encouragement in documenting the validations initially conducted for Ref.~\cite{Du:2023gnv}, which subsequently motivated this work's development. This work was supported in part by the Natural Sciences and Engineering Research Council of Canada, and in part by grant NSF PHY-1748958 to the Kavli Institute for Theoretical Physics (KITP). The author expresses appreciation for the hospitality during his stay at the KITP Program: {\it The Many Faces of Relativistic Fluid Dynamics}. Computations were made on the computers managed by the Ohio Supercomputer Center \cite{OhioSupercomputerCenter1987}.

\appendix
\renewcommand{\thefigure}{A\arabic{figure}}
\setcounter{figure}{0} 
\setcounter{equation}{0} 

\section{Performance of \isd{} at nonzero chemical potential}\label{app:is3d}

The \isd{} module \cite{McNelis:2019auj} is designed for sampling hadrons on a freeze-out hypersurface using the Cooper-Frye prescription \cite{Cooper:1974mv}. It includes multiple viscous correction models that characterize off-equilibrium effects of shear and bulk viscous stress, and baryon diffusion current. Additionally, it offers the ability to generate continuous distributions for particle species via the smooth Cooper-Frye formula without particle sampling. Both the sampling and continuous functionalities have been thoroughly tested and applied for heavy-ion studies. Extensive testing of \isd{} as a particle sampler under zero $\mu_B$ conditions has been conducted \cite{McNelis:2019auj,JETSCAPE:2020mzn,JETSCAPE:2020shq}. Furthermore, validation has been primarily focused on the continuous case for non-zero $\mu_B$ \cite{Du:2021zqz,Du:2021fyr,Du:2022oaw}, as well as the sampler case through comparison with the results obtained using \iss{} \cite{iss,Du:2023gnv}. 

\begin{figure}[!tbph]
\begin{center}
\includegraphics[width=0.8\linewidth]{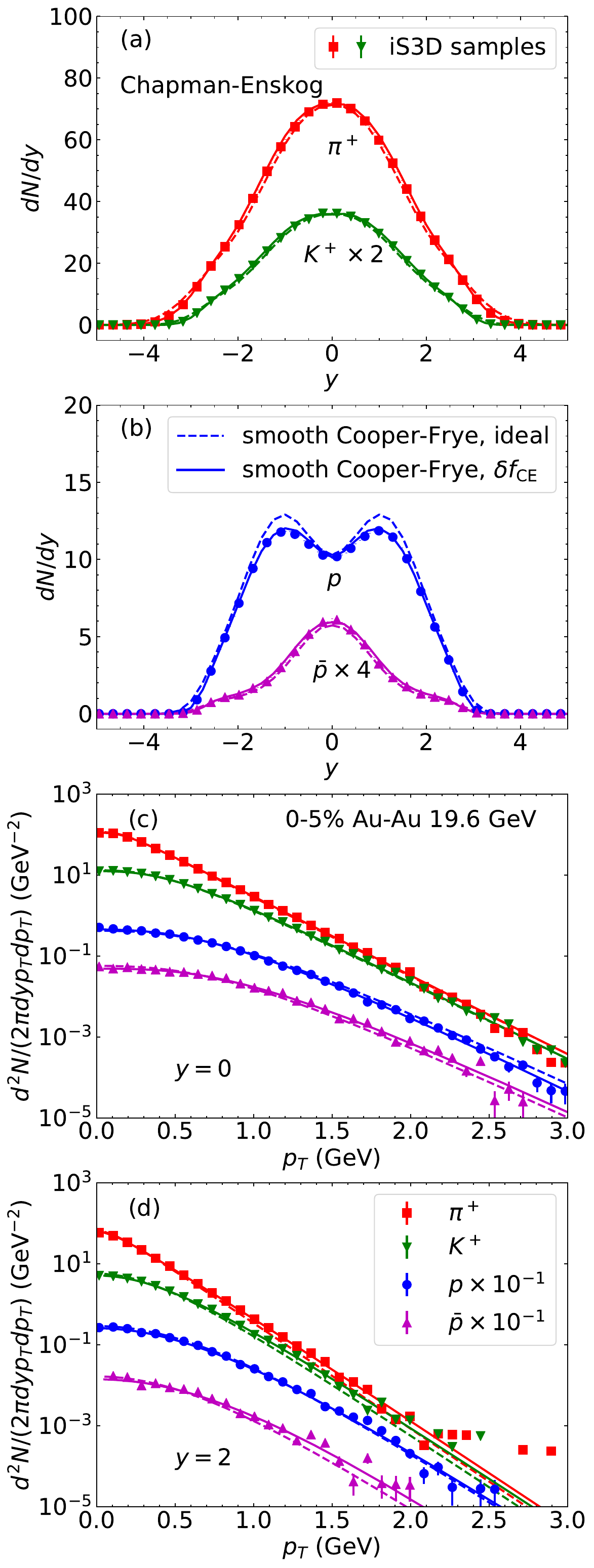}
\caption{%
    Validation of the \isd{} sampler conducted at non-zero net baryon chemical potential on a boost-non-invariant freeze-out hypersurface from 0-5\% Au+Au collisions at 19.6 GeV. The comparison involves distributions calculated via the smooth Cooper-Frye formula (solid lines) and the sampled particles (markers) for (a) rapidity densities of pions and kaons, (b) rapidity densities of protons and anti-protons, $p_T$-spectra of identified species (c) at $y=0$, and (d) at $y=2$. Viscous corrections from the Chapman-Enskog approximation for the shear stress tensor and baryon diffusion current are incorporated, while the smooth Cooper-Frye distributions without viscous corrections (``ideal'', dashed lines) are illustrated for comparison.
}
\label{fig:is3d_test_ce}
\end{center}
\end{figure}

This appendix aims to validate the \isd{} sampler at finite $\mu_B$ for viscous corrections modeled by the Chapman-Enskog approximation. Utilizing the freeze-out hypersurface from 0-5\% Au+Au collisions at $\snn=19.6$ GeV, I have obtained the rapidity distribution of identified particles (pions, kaons, and protons) along with their invariant momentum spectra at rapidities $y=0$ and $y=2$. These distributions are computed employing both the smooth Cooper-Frye distributions and the sampled particles. The comparisons are depicted in Fig.~\ref{fig:is3d_test_ce} for the Cooper-Frye formula incorporating viscous corrections of shear stress tensor and baryon diffusion current. Notably, the figure illustrates \isd{}'s proficient performance in accurately sampling various hadron species. Additionally, I present distributions without such viscous corrections, represented by dashed lines in Fig.~\ref{fig:is3d_test_ce}. It is important to highlight that the inclusion of viscous corrections visibly influences the distributions, indicating their significant effects on the results.

The \isd{} module is publicly maintained on GitHub, offering free accessibility for download \cite{is3d}. Additionally, I have integrated the \isd{} module into the {\sc iEBE-MUSIC} hybrid framework, combining \music{} with \isd{} and \urqmd{}. This comprehensive framework enables the simulation of heavy-ion collisions at BES energies. The simulations conducted in this study and Refs.~\cite{Du:2022yok,Du:2023gnv,Churchill:2023vpt,Churchill:2023zkk} are executed using this framework, which is also publicly available on GitHub for access \cite{iebe}.

\begin{figure}[!btp]
\begin{center}
\includegraphics[width= 0.9\linewidth]{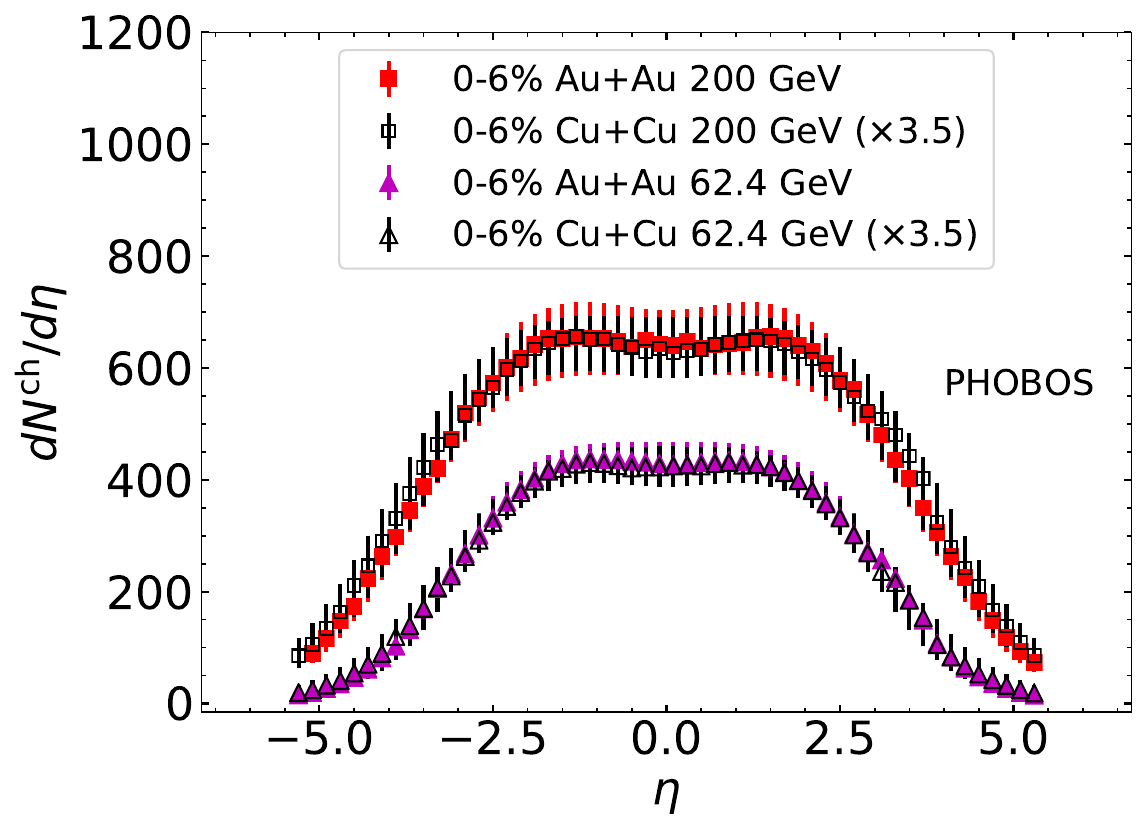}
\caption{%
    The charged particle multiplicity measured by the PHOBOS Collaboration for 0-6\% Au+Au collisions (solid markers) and Cu+Cu collisions (hollow markers) at beam energies of 200 GeV (square) and 62.4 GeV (triangle) \cite{PHOBOS:2010eyu}. The Cu+Cu collision multiplicity is scaled up using participant number ratios between Au+Au and Cu+Cu collisions.
    }
\label{fig:multi_scaling_AuCu}
\end{center}
\end{figure}

\section{Multisystem charged particle multiplicity}\label{app:aucu}

In Sec.~\ref{sec:reconst_27}, I employ participant number ratios, $N_\mathrm{part}^\mathrm{Au+Au}/N_\mathrm{part}^\mathrm{Cu+Cu}$, to scale up the charged particle multiplicity for Cu+Cu collisions at 22.4 GeV in order to estimate the multiplicity for Au+Au collisions at 27 GeV. Considering the constant nature of this ratio, this scaling does not alter the shape of the pseudo-rapidity distribution. It is reasonable to use a constant ratio due to the closely aligned beam rapidities at these energies, corresponding to similar distributions in rapidity or pseudo-rapidity. Here, I demonstrate the viability of utilizing Cu+Cu collision distributions to estimate those of Au+Au collisions by scaling up the former using participant number ratios, $N_\mathrm{part}^\mathrm{Au+Au}/N_\mathrm{part}^\mathrm{Cu+Cu}$. Figure~\ref{fig:multi_scaling_AuCu} visually demonstrates this validation, illustrating the remarkable agreement between the scaled-up multiplicity distributions for Cu+Cu collisions, based on participant number ratios, and those obtained from Au+Au collisions at corresponding beam energies. Thus, the agreement between the charged particle multiplicity distribution derived from this rescaling and the reconstructed distribution for 27 GeV in Fig.~\ref{fig:reconstruct_27}(a) indicates the efficacy of the distribution reconstruction method outlined in Sec.~\ref{sec:rap_reconst}.

\bibliography{ref}
\end{document}